\documentclass[a4paper,10pt]{article}
\usepackage[utf8]{inputenc}
\usepackage[a4paper, total={7in, 9in}]{geometry}

\usepackage{comment}
\usepackage{mathtools}
\usepackage{siunitx}
\usepackage{graphicx}
\usepackage{subcaption}
\usepackage{bbm}
\usepackage{pgfplots}
\usepackage{pgfplotstable}
\pgfplotsset{compat=1.3}
\usepackage{tikz}
\usepackage{authblk}
\usepgfplotslibrary{fillbetween}
\usetikzlibrary{patterns,calc,fit,plotmarks,tikzmark}

\usepackage{amsmath,amsbsy}
\usepackage[fleqn]{nccmath}
\usepackage{booktabs}
\usepackage{amsfonts}
\usepackage{amssymb}
\usepackage{caption}
\usepackage{cases}
\usepackage{comment}
\usepackage{graphicx}
\usepackage[colorlinks=true,bookmarksopen,bookmarksnumbered,linkcolor=black,citecolor=black,urlcolor=black]{hyperref}
\usepackage{mathtools}
\usepackage{moreverb}
\usepackage{multirow}

\usepackage{placeins}
\usepackage{tabularx}

\usepackage{multirow}
\usepackage{lmodern}
\usepackage{xparse}
\usepackage{colortbl}

\usepackage{glossaries}  

\newacronym{cad}{CAD}{computer-aided design}
\newacronym[longplural=degrees of freedom]{dof}{DoF}{degree of freedom}
\newacronym{fe}{FE}{finite element}
\newacronym{fem}{FEM}{finite element method}
\newacronym{fit}{FIT}{finite integration technique}
\newacronym{nurbs}{NURBS}{non-uniform rational B-splines}
\newacronym{pec}{PEC}{perfect electric conductor}
\newacronym{pmc}{PMC}{perfect magnetic conductor}
\newacronym{rf}{RF}{radio-frequency}
\newacronym{te}{TE}{transverse electric}
\newacronym{tm}{TM}{transverse magnetic}
\newacronym{tem}{TEM}{transverse electric and magnetic}
\newacronym[longplural=boundary conditions]{bc}{BC}{boundary condition}

\newacronym{uq}{UQ}{uncertainty quantification}
\newacronym{pml}{PML}{perfectly matched layers}
\newacronym{fdtd}{FDTD}{finite-difference time-domain}
\newacronym{dgtd}{DGTD}{discontinuous Galerkin time-domain}
\newacronym{mim}{MIM}{metal-insulator-metal}
\newacronym{rhs}{RHS}{right-hand side}
\newacronym{sqp}{SQP}{sequential quadratic programming}
\newacronym{lar}{LAR}{least angle regression}
\newacronym{td}{TD}{total degree}
\newacronym{tp}{TP}{tensor product}
\newacronym{hc}{HC}{hyperbolic cross}
\newacronym{lhs}{LHS}{latin hypercube sampling}
\newacronym{ls}{LS}{least squares}
\newacronym{qmc}{QMC}{quasi Monte Carlo}
\newacronym{mlmc}{MLMC}{multilevel Monte Carlo}
\newacronym{hf}{HF}{high-frequency}
\newacronym{megpc}{MEGPC}{multi-element generalized polynomial chaos}
\newacronym{gp}{GP}{Gaussian process}
\newacronym{ed}{ED}{experimental design}
\newacronym{rms}{RMS}{root mean square}
\newacronym{cv}{CV}{cross-validation}
\newacronym{es}{ES}{educated sampling}
\newacronym{iga}{IGA}{isogeometric analysis}
\newacronym{bem}{BEM}{boundary element method}
\newacronym{ibvp}{IBVP}{initial/boundary value problem}
\newacronym{em}{EM}{electromagnetic}
\newacronym{emf}{EMF}{electromagnetic field}
\newacronym{gpc}{gPC}{generalized polynomial chaos}
\newacronym{mc}{MC}{Monte Carlo}
\newacronym{pde}{PDE}{partial differential equation}
\newacronym{pdf}{PDF}{probability density function}
\newacronym[longplural={quantities of interest}]{qoi}{QoI}{quantity of interest}
\newacronym{rv}{RV}{random variable}
\newacronym{ae}{a.e.}{almost every}
\newacronym{ode}{ODE}{ordinary differential equation}
\newacronym{pce}{PCE}{polynomial chaos expansion}
\newacronym{mle}{MLE}{maximum likelihood estimation}

\newcommand{\mean}{\mathbb{E}}
\newcommand{\var}{\mathbb{V}}
\DeclareMathOperator*{\argmax}{\arg\!\max}

\definecolor{TUDa-0d}{cmyk/RGB/HTML}{0,0,0,.8/83,83,83/535353}
\definecolor{TUDa-0c}{cmyk/RGB/HTML}{0,0,0,.6/137,137,137/898989}
\definecolor{TUDa-0b}{cmyk/RGB/HTML}{0,0,0,.4/181,181,181/B5B5B5}
\definecolor{TUDa-0a}{cmyk/RGB/HTML}{0,0,0,.2/220,220,220/DCDCDC}

\definecolor{TUDa-1a}{cmyk/RGB/HTML}{.7,.4,0,0/93,133,195/5D85C3}
\definecolor{TUDa-2a}{cmyk/RGB/HTML}{0.8,.2,0,0/0,156,218/009CDA}
\definecolor{TUDa-3a}{cmyk/RGB/HTML}{0.7,0,.5,0/80,182,149/50B695}
\definecolor{TUDa-4a}{cmyk/RGB/HTML}{.4,0,.8,0/175,204,80/AFCC50}
\definecolor{TUDa-5a}{cmyk/RGB/HTML}{.2,0,.8,0/221,223,72/DDDF48}
\definecolor{TUDa-6a}{cmyk/RGB/HTML}{0,.1,.7,0/255,224,92/FFE05C}
\definecolor{TUDa-7a}{cmyk/RGB/HTML}{0,.3,.8,0/248,186,60/F8BA3C}
\definecolor{TUDa-8a}{cmyk/RGB/HTML}{0,.6,.8,0 /238,122,52/EE7A34}
\definecolor{TUDa-9a}{cmyk/RGB/HTML}{0,.8,.7,0/233,80,62/E9503E}
\definecolor{TUDa-10a}{cmyk/RGB/HTML}{.2,.9,0,0/201,48,142/C9308E}
\definecolor{TUDa-11a}{cmyk/RGB/HTML}{.6,.8,0,0/128,69,151/804597}
\definecolor{TUDa-1b}{cmyk/RGB/HTML}{1,.6,0,0/0,90,169/005AA9}
\definecolor{TUDa-2b}{cmyk/RGB/HTML}{1,.3,0,0/0,131,204/0083CC}
\definecolor{TUDa-3b}{cmyk/RGB/HTML}{1,0,.6,0/0,157,129/009D81}
\definecolor{TUDa-4b}{cmyk/RGB/HTML}{.5,0,1,0/153,192,0/99C000}
\definecolor{TUDa-5b}{cmyk/RGB/HTML}{.3,0,1,0/201,212,0/C9D400}
\definecolor{TUDa-6b}{cmyk/RGB/HTML}{0,.2,1,0/253,202,0/FDCA00}
\definecolor{TUDa-7b}{cmyk/RGB/HTML}{0,.4,1,0/245,163,0/F5A300}
\definecolor{TUDa-8b}{cmyk/RGB/HTML}{0,.7,1,0/236,101,0/EC6500}
\definecolor{TUDa-9b}{cmyk/RGB/HTML}{0,1,.9,0/230,0,26/E6001A}
\definecolor{TUDa-10b}{cmyk/RGB/HTML}{.4,1,0,0/166,0,132/A60084}
\definecolor{TUDa-11b}{cmyk/RGB/HTML}{.7,1,0,0/114,16,133/721085}
\definecolor{TUDa-1c}{cmyk/RGB/HTML}{1,.7,.2,0/0,78,138/004E8A}
\definecolor{TUDa-2c}{cmyk/RGB/HTML}{1,.5,.2,0/0,104,157/00689D}
\definecolor{TUDa-3c}{cmyk/RGB/HTML}{1,.2,.6,0/0,136,119/008877}
\definecolor{TUDa-4c}{cmyk/RGB/HTML}{.6,.1,1,0/127,171,22/7FAB16}
\definecolor{TUDa-5c}{cmyk/RGB/HTML}{.4,.1,1,0/177,189,0/B1BD00}
\definecolor{TUDa-6c}{cmyk/RGB/HTML}{.2,.3,1,0/215,172,0/D7AC00}
\definecolor{TUDa-7c}{cmyk/RGB/HTML}{.2,.5,1,0/210,135,0/D28700}
\definecolor{TUDa-8c}{cmyk/RGB/HTML}{.2,.8,1,0/204,76,3/CC4C03}
\definecolor{TUDa-9c}{cmyk/RGB/HTML}{.3,1,.9,0/185,15,34/B90F22}
\definecolor{TUDa-10c}{cmyk/RGB/HTML}{.5,1,.3,0/149,17,105/951169}
\definecolor{TUDa-11c}{cmyk/RGB/HTML}{.8,1,.2,0/97,28,115/611C73}
\definecolor{TUDa-1d}{cmyk/RGB/HTML}{1,.9,.3,0/36,53,114/243572}
\definecolor{TUDa-2d}{cmyk/RGB/HTML}{1,.7,.4,0/0,78,115/004E73}
\definecolor{TUDa-3d}{cmyk/RGB/HTML}{1,.4,.7,0/0,113,94/00715E}
\definecolor{TUDa-4d}{cmyk/RGB/HTML}{.7,.3,1,0/106,139,55/6A8B22}
\definecolor{TUDa-5d}{cmyk/RGB/HTML}{.5,.2,1,0/153,166,4/99A604}
\definecolor{TUDa-6d}{cmyk/RGB/HTML}{.4,.4,1,0/174,142,0/AE8E00}
\definecolor{TUDa-7d}{cmyk/RGB/HTML}{.3,.6,1,0/190,111,0/BE6F00}
\definecolor{TUDa-8d}{cmyk/RGB/HTML}{.4,.8,1,0/169,73,19/A94913}
\definecolor{TUDa-9d}{cmyk/RGB/HTML}{.5,1,.9,0/156,28,38/961C26}
\definecolor{TUDa-10d}{cmyk/RGB/HTML}{.7,1,.5,0/115,32,84/732054}
\definecolor{TUDa-11d}{cmyk/RGB/HTML}{.9,1,.3,0/76,34,106/4C226A}

\usepackage{tikz}
\usepackage{pgfplots}
\pgfplotsset{compat=newest}
\usepackage{subcaption}
\usepackage{siunitx}
\newcommand\BibTeX{{\rmfamily B\kern-.05em \textsc{i\kern-.025em b}\kern-.08em
T\kern-.1667em\lower.7ex\hbox{E}\kern-.125emX}}

\newcommand{\ra}[1]{\renewcommand{\arraystretch}{#1}}
\newcommand{\reviewag}[1]{\textcolor{black}{#1}}
\newcommand{\reviewur}[1]{\textcolor{black}{#1}}
\newcommand{\reviewdl}[1]{\textcolor{black}{#1}}

\providecommand{\keywords}[1]{\textbf{\textit{keywords---}} #1}
\begin{document}

\title{An \texorpdfstring{$hp$}{hp}-adaptive multi-element stochastic collocation method for surrogate modeling with information re-use}

\author{Armin Galetzka$^1$, Dimitrios Loukrezis$^{1,2,4}$, Niklas Georg$^{1,2,3}$, Herbert De Gersem$^{1,2}$, Ulrich R\"omer$^{3}$}

\date{
	\small{$^1$\emph{TU Darmstadt, Institute for Accelerator Science and Electromagnetic Fields (TEMF)} \\ \emph{Darmstadt, Germany}} \\ 
	\small{$^2$\emph{TU Darmstadt, Centre for Computational Engineering} \\ \emph{Darmstadt, Germany}} \\ 
	\small{$^3$\emph{TU Braunschweig, Institut für Dynamik und Schwingungen} \\ \emph{Braunschweig, Germany}} \\ 
	\small{$^4$\emph{Siemens AG, Technology} \\ \emph{Munich, Germany}} \\ 
}
\maketitle

\begin{abstract}This paper introduces an $hp$-adaptive multi-element stochastic collocation method, which additionally allows to re-use existing model evaluations during either $h$- or $p$-refinement. The collocation method is based on weighted Leja nodes. After $h$-refinement, local interpolations are stabilized by adding \reviewag{and sorting} Leja nodes on each newly created sub-element in a hierarchical manner. \reviewag{For $p$-refinement, the local polynomial approximations are based on total-degree or dimension-adaptive bases. The method is applied} in the context of forward and inverse uncertainty quantification to handle non-smooth or strongly localised response surfaces. \reviewdl{The performance of the proposed method is assessed in several test cases, also in comparison to competing methods.}	
		~\\~\\
		\noindent\keywords{$hp$-adaptivity; multi-element approximation; stochastic collocation; surrogate modeling; uncertainty quantification}
\end{abstract}

\section{Introduction}
\label{sec:introduction}
Prediction and optimization with computational models is now routinely carried out in many fields of science and engineering. For instance, reliable predictions require the propagation and quantification of uncertainties\cite{matthies2007quantifying} related to parameter calibration from noisy and limited data. The so-called non-intrusive approach is widely adopted, where computer codes are treated as black box functions from model parameters to \glspl{qoi}. In this way, restructuring and modifying complex software is avoided. 
Exemplarily, in forward \gls{uq} \cite{lee2009comparative}, also referred to as uncertainty propagation, non-intrusive methods evaluate the model at selected points of the input vector, which are generated according to the underlying probability distribution. 
Similarly, in inverse \gls{uq}\cite{bardsley2018computational}, repeated model evaluations are required to build the likelihood function at individual elements of a Markov chain. 

Even when large computational resources are available, such multi-query simulations with complex models present a widely acknowledged challenge. This fact has led to significant research efforts in the area of surrogate modeling \cite{jiang2020surrogate, koziel2013surrogate}, for example, by means of neural networks \cite{tripathy2018deep}, Gaussian processes \cite{sacks1989design}, or \glspl{pce} \cite{ghanem1990polynomial}.
Surrogate modeling can be carried out efficiently if the response function is sufficiently smooth and if the function varies moderately over the parameter space. However, there exist many examples in engineering and mechanics where the response is discontinuous or exhibits strong variations and sharp transitions \cite{le2004uncertainty}. In this case, most surrogate modeling techniques become too computationally expensive or altogether unreliable, thus, Monte Carlo methods are usually preferred. Nevertheless, an efficient sampling scheme is adapted to a specific simulation task and the computed model responses cannot be re-used in a different context. This is a major bottleneck of pure sampling approaches and a strong motivation to develop flexible and robust surrogate models, which can be applied even if the model response is non-smooth. 

\reviewur{Surrogate modeling for response surfaces with a reduced regularity typically involves some sort of local refinement. For instance, the spatially adaptive sparse grid method\cite{pflueger2010sgpp} allows for adaptive and hierarchical local refinement in subregions of the parameter domain. A combination of spatial and dimension-wise adaptivity has also been proposed\cite{obersteiner2021_adaptiveSparseGrids}. Alternatively, the family of multi-element methods\cite{chouvion2016development, halder2019adaptive, kawai2020multi, wan2005adaptive, foo2008multi} presents a class of approximation techniques which have been successfully applied in the \gls{uq} context for addressing non-smooth stochastic response functions. In a multi-element method, the parameter space is decomposed into subdomains referred to as \emph{elements}, where local polynomial basis functions are used. Here, on the one hand, one can opt for a discontinuity detection method, which first estimates a discontinuity and then splits elements accordingly. For example, the discontinuity detection can be based on spatially adaptive sparse grids and the element-wise restriction of sparse grid points is subsequently used for local polynomial approximation\cite{jakeman2013minimal}. A difficulty to be addressed here is the missing structure of the elements after splitting, which has previously been handled by the least orthogonal interpolation method\cite{jakeman2013minimal}. On the other hand, the element splitting can be based on hypercubes or simplices, which does not require any resolution of the discontinuity, however, applications in higher dimensions are more difficult to realize. We note that, local refinement can equally be applied in a Monte Carlo setting, e.g., using stratified sampling on simplices\cite{pettersson2022adaptive}.}

Here, we focus on the non-intrusive multi-element stochastic collocation method \reviewur{on hypercubes} for forward and inverse \gls{uq}. 
Although several proposals for multi-element collocation exist \cite{agarwal2009domain, foo2008multi, foo2010multi, fuchs2020simplex, jakeman2013minimal, ma2009adaptive, witteveen2013simplex}, a number of challenges remain to be addressed.
Most notably, computational efficiency considerations require to limit the number of collocation points, which is problematic since existing points do not necessarily fit to a collocation pattern once the subdomains in the parameter space are refined. Another challenge is to efficiently implement dimension adaptivity to avoid using large tensor grids in the parameter domain. \reviewag{Considering the worst-case where each parameter domain is split into two parts in each refinement step, the complexity of multi-element methods scales with $\mathcal{O}\left(2^N\right)$, where the exponent refers to the number of random inputs and the base to the subdomain splitting. This bottleneck, as mentioned above, is known to affect most multi-element methods suggested in the literature \cite{witteveen2012simplex, giovanis2019variance}. }
However, multi-element collocation methods can greatly benefit from $hp$-adaptivity concepts, which are well established in the finite element literature \cite{ainsworth1997posteriori}.
In this context, adaptive $h$-refinement, also referred to as $h$-adaptivity, refers to the adaptive decomposition of the computational domain into subdomains where local polynomial approximations are developed, while $p$-adaptivity refers to increasing the local polynomial degree within a subdomain. 
These concepts have already been successfully applied for \gls{uq} purposes in the context of the stochastic Galerkin method \cite{ge2022adaptive}. However, multi-element collocation methods so far rely on $h$-adaptivity only. \reviewag{In the best case, the $hp$-adaptive multi-element approach chooses the type of refinement that provides the highest convergence rate. That is, in elements where the function is locally analytic, $p$-refinement is carried out, leading to exponential convergence. 
Contrarily, $h$-refinement is performed in elements where the function shows reduced regularity, leading to algebraic convergence. For most practically relevant cases, this leads to large elements with a high regularity, where $p$-refinement is performed, and a fine grid around areas with reduced regularity, where $h$-refinement is preferred.}


In this work, we present a new $hp$-adaptive multi-element stochastic collocation method, which re-uses existing model evaluations during both $h$- and $p$-refinement. \reviewur{While re-using model data after splitting the parameter domain is easily possible for regression methods, this is not true for approaches relying on structured grids. Our method solves this problem through the use} of weighted Leja collocation points \cite{narayan2014adaptive}. A given sequence of collocation points can be stabilized by adding additional Leja points in a greedy way. \reviewag{Interpolation stability is further ensured by sorting the collocation points according to the Leja ordering.} Hence, on each element the collocation scheme can be stabilized by enriching the collocation set \reviewur{obtained after splitting}. This information re-use improves the overall efficiency of the method, which is demonstrated with several examples, also in comparison to existing approaches. \reviewur{Although our idea of stabilizing subsets of Leja sequences after multielement refinement is established here in the context of hypercubes, extensions to unstructured elements could be possible as well, for instance, with Leja sequences outlined in \cite{bos2010computing}}. 

\reviewag{Moreover, the weighted Leja nodes naturally handle arbitrary parameter distributions, which is particularly advantageous in the case of $h$-refinement with non-uniform distributions. }
\reviewag{For $p$-adaptivity, \gls{td} and dimension-adaptive\cite{chkifa2014high, loukrezis2019assessing} basis representations are utilized, but other choices of basis representations, e.g. based on Smolyak sparse grids, can be easily implemented. The dimension-adaptive algorithm relies on the nestedness and granularity of Leja points, which allows to increase the polynomial order in each dimension one-by-one, and exploits possible parameter anisotropies.}

Our approach contributes to the growing literature on adaptive surrogate models in the field of \gls{uq} and can be used both in forward and inverse \gls{uq} analysis, which we illustrate in the numerical examples. 
\reviewag{As a consequence of the underlying tensor product structure in case of $h$-refinement, we restrict ourselves to problems with low to moderate ($\mathcal{O}(10)$) dimensionality. 
In many practically relevant cases, high dimensional problems are reduced to moderate dimensions with the help of engineering experience and a-priori knowledge. Nevertheless, the remaining parameters may have a significant impact on the model response and can result in strong variations or discontinuities. The proposed approach is suitable to handle these types of problems.}

The remaining of this paper is structured as follows. 
In Section~\ref{sec:problem-setting}, we introduce the notation, the problem setting and some necessary preliminaries on stochastic collocation methods. Section~\ref{sec:leja-colloc-pce} is concerned with stochastic collocation on Leja grids, based either on hierarchical interpolation or on \gls{pce}. 
In Section~\ref{sec:me_leja}, we introduce our novel $hp$-adaptive multi-element stochastic method based on Leja grids, which additionally allows for re-using model evaluations after either $h$- or $p$-refinement. 
Numerical results showcasing the advantages of the proposed multi-element stochastic collocation method are reported in Section~\ref{sec:numerical_experiments}. 
Finally, conclusions are drawn in Section~\ref{sec:conclusion}. 

\section{\reviewur{Model Problem and Preliminaries}}
\label{sec:problem-setting}
In this section, we introduce the notation and the stochastic setting in particular. We also present the main ideas of stochastic collocation as a surrogate modeling approach for forward and inverse problems. Details of the collocation method are postponed until Section~\ref{sec:leja-colloc-pce}. \reviewur{Finally, we introduce a model problem based on a \gls{pde}.}

\subsection{\reviewur{Notation}}
\label{subsec:surrogate}
We consider a parameter-dependent model that receives as input a parameter vector and estimates an output, represented through the map 
\begin{equation}
    g: \mathbf{x} \mapsto g(\mathbf{x}).
    \label{eq:objective_function}
\end{equation}
In \eqref{eq:objective_function}, $\mathbf{x} \in \mathbb{R}^N$ denotes the input parameter vector and $g(\mathbf{x}) \in \mathbb{R}^{N_\mathrm{out}}$ the model evaluation for the given input parameters. 
In forward analyses, e.g., in the context of uncertainty propagation, $g(\mathbf{x})$ is typically called the \gls{qoi}. 
In inverse problem settings, $g(\mathbf{x})$ represents the map from parameters to observations or the likelihood function.
In cases where an evaluation $g(\mathbf{x})$ is computationally expensive, e.g., the model in question is a high-fidelity numerical solver, an inexpensive approximation $\tilde{g}(\mathbf{x}) \approx g(\mathbf{x})$ that can reliably replace the original model is often desirable. This is particularly true for multi-query tasks in \gls{uq}, optimization, or design space exploration.
We refer to such an approximation as a \emph{surrogate model}.

In the context of this work, the inputs $\mathbf{x}$ are assumed to be realizations of independent \glspl{rv} $X_n,\,n=1,2,\dots,N$, which are collected in the multivariate \gls{rv} $\mathbf{X}=(X_1,X_2,\dots,X_N)^\top$, also referred to as a random vector. 
The random vector $\mathbf{X}$ is defined on the probability space $(\Theta,\Sigma,P)$, where $\Theta$ denotes the sample space, $\Sigma$ the sigma algebra of events, and $P:\Sigma \rightarrow [0,1]$ the probability measure. 
We further introduce the image space $\Xi \subset \mathbb{R}^N$ such that $\mathbf{X}: \Theta \rightarrow \Xi$, equivalently, $\mathbf{X}(\theta) = \mathbf{x}$ with $\theta \in \Theta$ and $\mathbf{x} \in \Xi$, as well as the \gls{pdf} $\pi_{\mathbf{X}}(\mathbf{x}): \Xi \rightarrow \mathbb{R}_{\geq 0}$.
Note that the notation differentiates between a random vector $\mathbf{X}$ and a realization $\mathbf{x} = \mathbf{X}(\theta)$. 
Recalling that the individual \glspl{rv} $X_n$, $n=1,\dots,N$, are mutually independent, it holds that  $\pi_{\mathbf{X}}(\mathbf{x}) = \prod_{n=1}^N \pi_{X_n}(x_n)$ and $\Xi = \Xi_1 \times\cdots \times \Xi_N$, where $\pi_{X_n}(x_n)$ and $\Xi_n$ refer to the marginal (univariate) \glspl{pdf} and image spaces, respectively.
Then, the model output is a \gls{rv} dependent on the input random vector $\mathbf{X}$.
Note that the model itself remains purely deterministic, i.e., the uncertainty in the model output is caused only due to the random input parameters.
Accordingly, a \gls{rv} realization $\mathbf{x}$ corresponds to an  estimation $g(\mathbf{x})$ of fixed value. 

\subsection{\reviewur{Surrogate Modeling for Forward and Inverse Problems}}
\label{subsec:stochastic-collocation}
The term \emph{stochastic collocation} \cite{babuvska2010stochastic, xiu2005high} refers to a class of sampling-based methods where a model $g$ is evaluated (sampled) on a set of so-called \emph{collocation points} $\mathcal{X} = \left\{\mathbf{x}_j\right\}_{j=1}^J$. 
The pairs of collocation points and corresponding model evaluations $\left\{\mathbf{x}_i, g(\mathbf{x}_j)\right\}_{j=1}^J$ are then utilized to compute a surrogate model $\tilde{g}(\mathbf{x}) \approx g(\mathbf{x})$, which typically takes the form
\begin{equation}
\label{eq:polynomial-surrogate}
\tilde{g}(\mathbf{x}) = \sum_{i=1}^I \beta_i B_i(\mathbf{x}),
\end{equation}
where $\beta_i \in \mathbb{R}^{N_{\text{out}}}$ are coefficients, $B_i$ appropriately chosen basis functions, and $I$ the size of the basis. In a collocation approach, the coefficients are determined based on the condition 
\begin{equation}
    \label{eq:collocation-condition}
    g(\mathbf{x}_j) = \tilde{g}(\mathbf{x}_j), \ \forall j=1,\ldots,J,    
\end{equation}
where $I=J$, resulting in a linear system of equations for $\beta_i$. The well-known Lagrange basis even results in $\beta_i = g(\mathbf{x}_i)$. 
A major challenge for surrogate models in the form \eqref{eq:polynomial-surrogate} is the rapidly growing size of the basis when the input dimension is large. In this case, constructing a sparse basis has often proven to be useful \cite{barthelmann2000high, babuvska2010stochastic}. Even in case of sparse approximation, the construction of the polynomial basis relies on tensor product polynomials, hence, the one-dimensional case serves as the starting point. This approach is adopted in Section~\ref{sec:leja-colloc-pce}.

Before discussing details of the $hp$-adaptive collocation approach, we briefly outline the role of surrogate models in inverse and forward UQ. In forward UQ, the goal is to compute moments or a distribution function of the \gls{qoi}. This can be achieved, either by sampling the inexpensive surrogate model, or by choosing a suitable basis from which the sought quantities can be readily obtained. The polynomial chaos basis, for instance, allows to infer the moments directly from the basis coefficients without involving any additional sampling routine\cite{ghanem1990polynomial, sudret2008global}. In inverse UQ, an important object is the likelihood function which, in the common case of Gaussian noise, reads 
\begin{equation}
    \label{eq:negLogLikelihood}
    L(\mathbf{x}|\mathbf{b}) = \mathrm{exp}\left(-\frac{1}{2} \| \mathbf{b} - \mathcal{G}(\mathbf{x}) \|_{\boldsymbol{\Sigma}^{-1}} \right),
\end{equation}
where $\mathbf{b} \in \mathbb{R}^D$ denotes the observation vector and $\boldsymbol{\Sigma}$ the noise covariance matrix. Then, $\mathcal{G}:\mathbb{R}^N \to \mathbb{R}^D$ represents the nonlinear relationship between input parameters and observations. In \eqref{eq:negLogLikelihood}, the model-data mismatch is measured in the discrete Euclidean norm, weighted with the inverse covariance matrix. When running a Markov chain Monte Carlo analysis, the likelihood function and hence, the model included in $\mathcal{G}$, needs to be evaluated for each entry of the chain. To reduce the large computational workload, a surrogate model $\tilde{g}(\mathbf{x}) \approx g(\mathbf{x}) = \mathcal{G}(\mathbf{x})$ can be employed \cite{marzouk2009stochastic}. A more recent approach constructs a surrogate for the likelihood as $\tilde{g}(\mathbf{x}) \approx g(\mathbf{x}) = L(\mathbf{x}|\mathbf{b})$, which allows to avoid any sampling-based analysis \cite{wagner2021bayesian}. However, in the large data - small noise regime, the likelihood function is highly concentrated and global surrogate modeling is difficult. \reviewur{Adaptive\cite{schillings2013sparse} and possibly multilevel\cite{farcas2020multilevel} approaches are a popular remedy in this case. Another possibility} is to use the spectral stochastic embedding (SSE) method \cite{wagner2021bayesian}. \reviewur{Other approaches are based on transport maps \cite{eigel2022low,parno2018transport}.}
We will show in Section \ref{sec:numerical_experiments}, that our multi-element stochastic collocation method is also capable of handling such highly concentrated likelihood functions.  


\subsection{\reviewur{A PDE-based Model Problem}}
\label{sec:pde_model_problem}
\reviewur{For the sake of concreteness, we consider the Helmholtz equation with a stochastic inhomogeneous refractive index as a model problem. However, the formulation derived in this subsection will be general enough to cover other problems as well. Let $D = [0,1]^2$ denote the computational domain with boundaries}
\[
\reviewur{\Gamma_1 = \{ (r_1,r_2) \ | \ r_1=0, r_2 \in [0,1]\}, \quad \Gamma_2 = \{ (r_1,r_2) \ | \ r_1=1, r_2 \in [0.3,0.7]\}, \quad \Gamma_3 = \partial D \setminus \left(\Gamma_1 \cup \Gamma_2\right).}
\]
\reviewur{Let $\partial_\nu$ denote the derivative with respect to the outgoing normal vector. The strong form of the problem reads}
\begin{subequations}
\begin{alignat}{3}
    \reviewur{\Delta u(\mathbf{x};\mathbf{r}) + k^2 n(\mathbf{x};\mathbf{r}) u(\mathbf{x};\mathbf{r})} &= 0 \quad &&\reviewur{\mathbf{r}\in D}, \\
    \reviewur{\partial_\nu u(\mathbf{x};\mathbf{r})} &= \reviewur{g(\mathbf{r})} \quad &&\reviewur{\mathbf{r} \in \Gamma_1},\\
    \reviewur{\partial_\nu u(\mathbf{x};\mathbf{r})} &= \reviewur{(i \beta - \alpha) u(\mathbf{x};\mathbf{r})} \quad &&\reviewur{\mathbf{r} \in \Gamma_2}, \\
    \reviewur{\partial_\nu u(\mathbf{x};\mathbf{r})} &= \reviewur{0} \quad &&\reviewur{\mathbf{r} \in \Gamma_3},
\end{alignat}%
\end{subequations}%
\reviewur{for $\pi_{\mathbf{X}}$-almost all $\mathbf{x} \in \Xi$. The uncertainty in the index $n$ is modelled with a log-Karhunen-Lo\`eve expansion of the form}
\begin{equation}
\label{eq:log-KLE}
    \reviewur{n(\mathbf{x};\mathbf{r}) = \exp \left( n_0(\mathbf{r}) + \sum_{j=1}^N x_j \Psi_j(\mathbf{r}) \right)}.
\end{equation}
\reviewur{Here, the modes $\Psi_j(\mathbf{r}) = \sqrt{\lambda_j} \varphi_j(\mathbf{r})$ are obtained from the squared exponential kernel with correlation length $l=0.05$ and standard deviation $\sigma=1$. In the numerical experiments we set $n_0=0$, see Section~\ref{subsubsec:Helmholtz}.}

\reviewur{We consider a $\Xi$-strong/$D$-weak form based on $H^1(D)$, where we seek for $u(\mathbf{x}) \in H^1(D)$ $\pi_{\mathbf{X}}$-almost everywhere in $\Xi$ subject to}
\begin{equation}
    \label{eq:Helmholtz_weak}
    \reviewur{\underbrace{\int_{D} \nabla u(\mathbf{x}) \cdot \nabla v^* \ \mathrm{d} r - k^2 \int_{D} n(\mathbf{x}) u(\mathbf{x}) v^* \ \mathrm{d} r - \int_{\Gamma_2} (i \beta - \alpha) u(\mathbf{x}) v^* \mathrm{d} r}_{=a(\mathbf{x};u,v)}= \underbrace{\int_{\Gamma_1} g v^* \mathrm{d} r}_{=l(v)}, \quad \forall v \in H^1(D),}
\end{equation}
\reviewur{where $^*$ refers to the complex conjugate and where $u(\mathbf{x})$ is short for $u(\mathbf{x};\cdot)$. We apply a standard, lowest order, piecewise continuous finite element method on a regular grid with maximum mesh size $h$, to compute a numerical solution $u_h(\mathbf{x}) \in V_h\subset H^1(D)$, subject to}
\begin{equation}
    \label{eq:Helmholtz_primal}
    \reviewur{a(\mathbf{x};u_h,v_h) = l(v_h), \quad \forall v_h \in V_h}.
\end{equation}
\reviewur{The \gls{qoi} is here obtained as a linear functional of the solution, i.e.,
\begin{equation}
    g(\mathbf{x}) = q(u_h(\mathbf{x})), \quad q \in V_h^*.
\end{equation}
The \gls{pde}-background of the problem provides more structure, which we exploit to derive an error indicator for the $hp$-adaptive method. 
}

\section{Stochastic collocation on Leja grids}
\label{sec:leja-colloc-pce}
Next, we elaborate on the concept of hierarchical Leja collocation in one- and multiple dimensions and its relation to polynomial chaos surrogates. \reviewdl{These topics have been previously presented in the literature in detail, nevertheless, we shortly re-iterate the relevant concepts for the ease of the reader.}

\subsection{Leja sequences}
\label{subsec:leja-sec}
In their original form, Leja sequences \cite{leja1957sur} are sequences of points $\left\{x_i\right\}_{i \geq 0}$, $x_i \in \left[-1,1\right]$, $\forall i \geq 0$, where the initial point $x_0$ is chosen arbitrarily within $\left[-1,1\right]$ and the remaining points are computed by solving the optimization problem
\begin{equation}
 \label{eq:leja_unweighted}
 x_i = \argmax_{x \in \left[-1,1\right]} \prod_{k=0}^{i-1} \left|x-x_k\right|.
\end{equation}
A point sequence produced by formula \eqref{eq:leja_unweighted} is an \emph{unweighted} Leja sequence.
\emph{Weighted} Leja sequences are constructed by incorporating a continuous and positive weight function in the definition of Leja sequences \cite{narayan2014adaptive}.
Assuming that the weight function is a \gls{pdf} $\pi_X(x): \Xi \rightarrow \mathbb{R_{\geq 0}}$, the points of the corresponding weighted Leja sequence are given as
\begin{equation}
 \label{eq:leja_weighted}
 x_i = \argmax_{x \in \Xi} \sqrt{\pi_X(x)}\prod_{k=0}^{i-1} \left|x-x_i\right|,
\end{equation}
where again the initial Leja point $x_0$ is chosen arbitrarily within the image space $\Xi$.

Leja points are known to perform well when employed as interpolation or quadrature nodes \cite{bos2015application, carnicer2019central, carnicer2017optimal, loukrezis2019assessing, reichel1990newton}. 
With respect to interpolation in particular, the Lebesgue constant of Leja sequence based interpolation grids is known to grow subexponentially \cite{calvi2011lebesgue, jantsch2019lebesgue, taylor2008lebesgue}, thus resulting in stable interpolations.
Additionally, interpolation and quadrature grids with respect to any continuous \gls{pdf} can be constructed by using weighted Leja sequences \cite{farcas2020multilevel, jakeman2019polynomial, loukrezis2020approximation, vandenbos2019bayesian}.
Moreover, due to the fact that Leja sequences are by definition nested, i.e., $\left\{x_i\right\}_{i=0}^j \subset \left\{x_i\right\}_{i=0}^{j+1}$, they allow for re-using readily available Leja points and model evaluations on those points in case the sequence is further expanded. 
Due to the nestedness property, Leja points are natural candidates for constructing sparse interpolation or quadrature grids  \cite{chkifa2014high, farcas2020sensitivity, georg2020enhanced, nobile2015comparison, schillings2013sparse}.
Note that nested interpolation and quadrature grids can be obtained with other types of nodes, such as Clenshaw-Curtis \cite{clenshaw1960method}. 
However, Leja sequences have the additional attractive feature of retaining the nestedness property even if a single point is added to the sequence, thus allowing for arbitrary grid granularity. 
Nestedness and grid granularity become additionally important \reviewdl{when} considering efficient $h$-refinement, as discussed in Section~\ref{sec:me_leja}.

\subsection{Hierarchical interpolation on Leja grids}
\label{subsection:leja-colloc}
In interpolation based stochastic collocation methods \cite{babuvska2010stochastic, barthelmann2000high, xiu2005high}, the surrogate model $\tilde{g}(\mathbf{x}) \approx g(\mathbf{x})$ is a global polynomial approximation computed by means of interpolation.
For simplicity, in the following we assume a scalar model output $g(\mathbf{x}) \in \mathbb{R}$, however, the extension to vector-valued outputs is straightforward.

Considering a model $g(x)$ dependent on a single parameter and an interpolation grid $\mathcal{X}_j = \left\{x_i\right\}_{i=0}^j$, a Lagrange interpolation based approximation is given as
\begin{equation}
\label{eq:lagrange-interpolation-1d}
\tilde{g}(x) = \mathcal{I}_j[g](x) = \sum_{i=0}^{j} g\left(x_i\right) l_i^j(x),
\end{equation}
where $l_i^j$ are Lagrange polynomials of degree $j$, defined as
\begin{align}
\label{eq:lagrange-polynomial-1d}
l_i^j(x) = \prod_{k=0,k\neq i}^{j} \frac{x - x_k}{x_i - x_k}, \qquad l^0 = 1.
\end{align}
Assuming that the interpolation grid $\mathcal{X}_j$ coincides with a Leja sequence, a sequence of nested grids can be defined, such that $\mathcal{X}_0 =\left\{x_0\right\} \subset \mathcal{X}_1 = \left\{x_0, x_1\right\} \subset \cdots \subset \mathcal{X}_j = \left\{x_0, x_1, \dots, x_j\right\}$, where each grid defines an interpolation operator $\mathcal{I}_i[g](x)$, $i=0,1,\dots,j$.
Then, replacing the Lagrange polynomials with the hierarchical polynomials \cite{chkifa2014high}
\begin{align}
\label{eq:hierarchical-polynomial-1d}
h^i(x) = \prod_{k=0}^{i-1} \frac{x-x_k}{x_i - x_k}, \qquad h^0(x)=1,
\end{align}
of polynomial degree $i$, $i=0,1,\dots,j$, a sequential interpolation formula can be derived, such that
\begin{align}
\label{eq:hierarchical-interpolation-1d}
\tilde{g}(x) = \mathcal{I}_j[g](x) = \sum_{i=0}^j s_i h^i(x) = \sum_{i=0}^j \left(g\left(x_i\right) - \mathcal{I}_{i-1}[g]\left(x_i\right)\right)
h^i(x),
\end{align}
where the coefficients $s_i$ are called the hierarchical surpluses\cite{gerstner2003dimension}. 
Note that $\mathcal{I}_{-1}$ is a null operator, i.e., $\mathcal{I}_{-1}[g](x)=0$, accordingly, $s_0 = g(x_0)$. 
We shall refer to the interpolation format \eqref{eq:hierarchical-interpolation-1d} as \emph{hierarchical} interpolation.

The interpolation formats \eqref{eq:lagrange-interpolation-1d} and \eqref{eq:hierarchical-interpolation-1d} are in fact equivalent, as the interpolating polynomial is unique for a given interpolation grid irrespective of the choice of the polynomial basis \cite{gander2005change}.
The hierarchical representation \eqref{eq:hierarchical-interpolation-1d} offers the advantage that, if the Leja sequence based interpolation grid is expanded with new nodes, the already computed basis polynomials remain unchanged. 
Accordingly, each hierarchical polynomial $h^i$, $i=0,1,\dots,j$, has a degree equal to $i$ which is unique. 
Irrespective of the choice of basis polynomials, a crucial advantage of using Leja sequences as interpolation grids is that the readily available model evaluations $g(x_i)$, $x_i \in \mathcal{X}_{j}$, $i=0,1,\dots,j$, that are used to construct the interpolation $\mathcal{I}_{j}[g](x)$, can be re-used for the interpolation $\mathcal{I}_{j+1}[g](x)$, due to the fact that $\mathcal{X}_j \subset \mathcal{X}_{j+1}$. Accordingly, the model needs to be evaluated only on the new Leja point $x_{j+1}$, where $\left\{x_{j+1}\right\} = \mathcal{X}_{j+1} \setminus \mathcal{X}_{j}$.

Moving to an $N$-variate model $g(\mathbf{x})$, we introduce a multi-index notation such that a multi-index $\mathbf{j} = (j_1, \dots, j_N)$ can be uniquely associated with the multivariate Leja node $\mathbf{x}_{\mathbf{j}} = \left(x_{1,j_1}, \dots, x_{N,j_N}\right)$, the multivariate hierarchical polynomial
\begin{equation}
\label{eq:hierarchical-polynomial-Nd}
H_{\mathbf{j}}(\mathbf{x}) = \prod_{n=1}^N h_n^{j_n}(x_n),
\end{equation}
and the tensor grid
\begin{equation}
\mathcal{X}_{\mathbf{j}} = \mathcal{X}_{1,j_1} \times \mathcal{X}_{2,j_2} \times \cdots \times \mathcal{X}_{N,j_N},
\end{equation}
where each univariate grid $\mathcal{X}_{n,j_n} = \left\{x_{n,0}, x_{n,1}, \dots,x_{n,j_n}\right\}$ is assumed to be a Leja sequence.
A multivariate hierarchical interpolation scheme can then be derived as follows.
We consider a downward closed multi-index set $\Lambda$, such that
\begin{equation}
\label{eq:downward-closed}
\forall\mathbf{j} \in \Lambda \Rightarrow \mathbf{j} - \mathbf{e}_n \in \Lambda, \forall n=1,\dots,N,
\end{equation}
where $\mathbf{e}_n$ denotes the unit vector in the $n$th dimension. 
The associated interpolation grid is given by
\begin{equation}
\mathcal{X}_{\Lambda} = \bigcup_{\mathbf{j} \in \Lambda} \mathcal{X}_{\mathbf{j}}.
\end{equation}
Since each multi-index $\mathbf{j} \in \Lambda$ uniquely defines a multivariate Leja node $\mathbf{x}_{\mathbf{j}} \in \mathcal{X}_{\Lambda}$, it holds that $\#\Lambda = \#\mathcal{X}_{\Lambda}$, where $\#$ denotes the cardinality of a set.
Then, a sequence of nested, downward closed multi-index sets $\left(\Lambda_w\right)_{w=1}^W$, $\#\Lambda=W$, can be defined, such that $\Lambda_w = \left\{\mathbf{j}_1, \mathbf{j}_2, \dots, \mathbf{j}_w\right\}$, $\Lambda_{w-1} \subset \Lambda_w$ ($\Lambda_0 = \emptyset$), $\#\Lambda_w=w$, and $\Lambda_w\setminus\Lambda_{w-1} = \left\{\mathbf{j}_w\right\}$. 
This also implies that $\Lambda_1 = \left\{\mathbf{j}_1 = \mathbf{0} = \left(0,0,\dots,0\right)\right\}$.
A corresponding grid sequence $\left(\mathcal{X}_{\Lambda_w}\right)_{w=1}^W$ is similarly defined, where $\mathcal{X}_{\Lambda_w} = \left\{\mathbf{x}_{\mathbf{j}_1}, \mathbf{x}_{\mathbf{j}_2}, \dots, \mathbf{x}_{\mathbf{j}_w}\right\}$, $\mathcal{X}_{\Lambda_{w-1}} \subset \mathcal{X}_{\Lambda_w}$ ($\mathcal{X}_{\Lambda_0} = \emptyset$), $\#\mathcal{X}_{\Lambda_w} = w$, and $\mathcal{X}_{\Lambda_{w}} \setminus \mathcal{X}_{\Lambda_{w-1}} = \left\{\mathbf{x}_{\mathbf{j}_w}\right\}$, i.e., a single Leja node is added to the next grid in sequence, as in the univariate case.
Simplifying the notation to $\mathbf{x}_{\mathbf{j}_w} = \mathbf{x}_w$ and $H_{\mathbf{j}_w}(\mathbf{x}) = H_w(\mathbf{x})$, the multivariate hierarchical interpolation reads
\begin{align}
\label{eq:hierarchical-interpolation-Nd}
\tilde{g}(\mathbf{x}) = \mathcal{I}_{\Lambda}[g](\mathbf{x}) 
= \sum_{\mathbf{j} \in \Lambda} s_{\mathbf{j}} H_{\mathbf{j}}(\mathbf{x})
= \sum_{w=1}^W s_w H_w(\mathbf{x})
= \sum_{w=1}^{W} \left(g\left(\mathbf{x}_w\right) - \mathcal{I}_{\Lambda_{w-1}}[g]\left(\mathbf{x}_w\right)\right) H_w(\mathbf{x}) ,
\end{align}
where $s_1 = s_{\mathbf{0}} = g(\mathbf{x}_1) = g(\mathbf{x}_{\mathbf{0}})$, i.e., $\mathcal{I}_{\Lambda_0}$ is a null operator, similar to the univariate case.

\reviewdl{ 
The hierarchical Leja interpolation \eqref{eq:hierarchical-interpolation-Nd} can be constructed by means of a so-called dimension-adaptive algorithm, which sequentially adds terms of high impact to the polynomial approximation, while neglecting non-influential terms.
This algorithm was first suggested for adaptive, sparse quadrature \cite{gerstner2003dimension}. 
Similar algorithms have been employed for Leja based, dimension-adaptive interpolation \cite{narayan2014adaptive, chkifa2014high, loukrezis2019assessing}. 
In brief, the dimension-adaptive Leja interpolation algorithm consists of the following steps:
\begin{enumerate}
\item A hierarchical interpolation based on a downward closed multi-index set $\Lambda$ is assumed to exist, along with the corresponding Leja interpolation grid $\mathcal{X}_{\Lambda}$. Otherwise, the algorithm is initialized with the zero multi-index, such that $\Lambda = \left\{\mathbf{0}\right\}$. This set is referred to as the \emph{active} set.
\item The set of \emph{admissible} multi-indices is computed, denoted as $\Lambda^{\text{adm}}$, such that $\Lambda \cup \Lambda^{\text{adm}}$ is downward closed. The corresponding admissible Leja nodes are also computed.
\item The hierarchical surplus corresponding to each admissible multi-index is computed. 
\item The admissible multi-index with the maximum hierarchical surplus in absolute value is added to the active multi-index set. The corresponding Leja node is added to the interpolation grid.
\item The steps (2) - (5) are repeated until the desired approximation accuracy or the maximum allowed number of model evaluations is reached.
\end{enumerate}
After the termination of the algorithm, the final interpolation is constructed using both the active and the admissible multi-indices, such that no Leja nodes and corresponding model evaluations remain unused.
}

\reviewdl{
Alternatively, Leja based interpolations can be constructed using interpolation grids with an a priori defined structure, for example, tensor product, \glsfirst{td}, hyperbolic cross, or Smolyak sparse grids\cite{babuvska2010stochastic}. In this work, aside from the dimension-adaptive algorithm described above,  \gls{td} grids are also employed, in which case the multi-index set is given as
\begin{equation}
\Lambda^{\text{TD}} = \left\{\mathbf{j} : \lVert \mathbf{j} \rVert_1 \leq P, P \in \mathbb{Z}_{\geq 0} \right\}.
\end{equation}
These two choices regarding the employed multi-index set are mainly motivated by the fact that they greatly simplify the transformation to a \gls{pce} basis, see Section~\ref{subsec:leja-pce}.
}

\subsection{Polynomial chaos expansion on Leja grids}
\label{subsec:leja-pce}
Similar to the stochastic collocation method presented in Section~\ref{subsection:leja-colloc}, the \gls{pce} method \cite{ghanem2003stochastic, xiu2002wiener} also produces a global polynomial approximation $\tilde{g}(\mathbf{x}) \approx g(\mathbf{x})$, however, instead of Lagrange or hierarchical polynomials, the polynomial basis consists of orthonormal polynomials with respect to the \gls{pdf} of the input parameters.
Again, in the following, we consider a scalar model output $g(\mathbf{x}) \in \mathbb{R}$ for simplicity.
The method can be applied to address vector-valued model outputs as well.

\reviewdl{
The \gls{pce} is given as
\begin{equation}
	\tilde{g}(\mathbf{x}) = \sum_{\mathbf{p} \in \Lambda} c_{\mathbf{p}} \Psi_{\mathbf{p}} (\mathbf{x}),
	\label{eq:gpc_ansatz}
\end{equation}
where $c_{\mathbf{p}}$ are scalar coefficients and $\Psi_{\mathbf{p}}(\mathbf{x}) = \prod_{n=1}^N \psi_n^{p_n}(x_n)$ are multivariate polynomials that satisfy the orthonormality condition
\begin{equation}
\label{eq:orthonormality-Nd}
\mean\left[\Psi_{\mathbf{p}} \Psi_{\mathbf{q}}\right] = \int_\Xi \Psi_{\mathbf{p}}(\mathbf{x}) \Psi_{\mathbf{q}}(\mathbf{x})\pi_{\mathbf{X}}(\mathbf{x}) \,\mathrm{d}\mathbf{x} = \delta_{\mathbf{p}\mathbf{q}},
\end{equation}
where $\mean[\cdot]$ denotes the expectation operator, $\delta_{\mathbf{p}\mathbf{q}} = \prod_{n=1}^N \delta_{p_n q_n}$, and $\delta_{ij}$ is the Kronecker delta.
The multi-index set $\Lambda$ now comprises multi-indices corresponding to the polynomial degrees of the \gls{pce} polynomials, i.e.,   
$\mathbf{p} = \left(p_1, \dots,p_N\right)$.
}

\reviewdl{
Note that, assuming a readily available Lagrange or hierarchical interpolation, an equivalent \gls{pce} can be computed with a basis transformation to orthonormal polynomials \cite{buzzard2012global, buzzard2013efficient, porta2014inverse}. 
As previously noted in Section~\ref{subsection:leja-colloc}, in this work, such transformations take place using Leja interpolations based either on \glsfirst{td} multi-index sets\cite{babuvska2010stochastic}, which have an a priori defined structure, or multi-index sets constructed with a dimension-adaptive algorithm. 
In both cases, there exists a one-to-one relation between the hierarchical and orthonormal polynomials in terms of their degrees. Therefore, the interpolation and \gls{pce} multi-index sets are identical and the basis transformation is greatly simplified\cite{loukrezis2019adaptive}.
}

\reviewdl{
An important aspect of the \gls{pce} is that it provides \gls{uq} metrics regarding the approximated model output, which are obtained with negligible computational cost by simply post-processing the \gls{pce}'s terms. 
In particular, exploiting the orthonormality property of the basis polynomials, the mean value and the variance of the output can be estimated as
\begin{align}
\mu &= c_{\mathbf{0}}, \label{eq:pce-mean} \\
\sigma^2 &= \sum_{\mathbf{p} \in \Lambda\setminus\left\{\mathbf{0}\right\}}  c_{\mathbf{p}}^2. \label{eq:pce-variance}
\end{align}
Note that each \gls{pce} coefficient $c_{\mathbf{p}}$, $\mathbf{p}\in \Lambda \setminus \left\{\mathbf{0}\right\}$, corresponds to a partial variance that contributes to the total variance of the model output. 
Based on the known connection between variance based sensitivity analysis metrics and the \gls{pce} method \cite{crestaux2009polynomial, sudret2008global}, all \gls{pce} coefficients except for $c_{\mathbf{0}}$ can be interpreted as sensitivity indicators \cite{farcas2020sensitivity, loukrezis2020robust}.
}

\section{Multi-element stochastic collocation with adaptive refinement}
\label{sec:me_leja}
Global polynomial approximations face major difficulties in terms of accuracy and convergence when applied to problems with discontinuities in the parameter space. Additionally, steep gradients or large variations in the objective function also hinder the convergence of global approximation approaches.
To address this bottleneck, several \emph{multi-element} approximation techniques have been proposed \cite{agarwal2009domain, chouvion2016development, jakeman2013minimal, foo2008multi, foo2010multi, wan2006multi, wan2005adaptive}, where the parameter space is decomposed into subspaces of improved regularity. 
Local polynomial approximations are developed in the subspaces, which are subsequently combined to provide a global surrogate model. 
Borrowing terminology from the \gls{fem} literature \cite{burg2011convergence, demkowicz2006computing, dorfler2007convergence, houston2002discontinuous, mitchell2014comparison}, we call $h$-refinement the refinement of the parameter domain decomposition with additional subdomains and $p$-refinement the refinement of a local polynomial approximation with additional terms. 
In the following, we build upon the Leja based hierarchical interpolation and \gls{pce} methods presented in Section~\ref{sec:leja-colloc-pce} and develop a multi-element polynomial approximation method which refines adaptively both the parameter domain decomposition and the polynomial approximations according to the particularities of the problem at hand.

\subsection{Multi-element hierarchical interpolation and polynomial chaos expansion}
\label{sec:decomp}
\reviewag{Following the work of Wan and Karniadakis \cite{wan2005adaptive},} let the image space of the random vector $\mathbf{X}=\left(X_1, \dots,X_N\right)$ be given as $\Xi=[a_1,b_1]\times[a_2,b_2] \times \dots \times [a_N,b_N] $, where $a_n, b_n \in \mathbb{R}$, $n=1,\dots,N$, with extension to infinity.
The image space is decomposed into $K$ non-overlapping hypercubes, subsequently referred to as \emph{elements}, such that $\Xi=\bigcup_{k=1}^K d_k$, where each element $d_k$ is defined as $d_k = d_1^{(k)} \times d_2^{(k)} \times \dots \times d_N^{(k)}$ with \reviewag{$d_n^{(k)}=[a_n^{(k)},b_n^{(k)})$ if $b_n^{(k)} \neq b_n$, respectively $d_n^{(k)}=[a_n^{(k)},b_n^{(k)}]$ if $b_n^{(k)} = b_n$}.

We proceed by introducing conditional expectations and local densities associated to the partition $\{d_k\}$. To that end, let $\mathbb{1}_k(\mathbf{x})$ denote the indicator function defined as
\begin{equation}
    \mathbb{1}_k(\mathbf{x}) = \left\{\begin{array}{ll} 1, & \mathbf{x}\in d_k, \\
         0, & \mathrm{else}.\end{array}\right.
\end{equation}
We write $Y_k= \mathbb{1}_k(\mathbf{X})$ in the following. Then, 
\begin{equation}
    \mathbb{E}[\mathbf{X}|Y_k=1] = \frac{\mathbb{E}[\mathbf{X} Y_k]}{P(Y_k=1)}, \ k=1,\ldots,K.
\end{equation}
We then use the definition of a PDF via an expected value 
\begin{equation}
    \pi_X(x) = \mathbb{E}[\delta(X-x)],
\end{equation}
where $\delta$ denotes the Dirac delta distribution, to obtain conditional PDFs for the partition. In particular, we obtain 
\begin{align}
    \pi_{\mathbf{X}}(\mathbf{x}|Y_k=1) &= \frac{\mathbb{E}[\delta(\mathbf{X}-\mathbf{x})Y_k]}{P(Y_k=1)}, \\ 
    \pi_k(\mathbf{x}^{(k)}) &= \pi_{\mathbf{X}}(\mathbf{x}^{(k)}|Y_k=1) = \frac{\pi_\mathbf{X}\left(\mathbf{x}^{(k)}\right)}{\varrho_k}, \quad \text{for } \mathbf{x}= \mathbf{x}^{(k)} \in d_k.
\end{align}
The normalization factor $\varrho_k = \mathrm{Pr}(Y_k = 1)$ is determined as
\begin{equation}
    \varrho_k = \prod_{n=1}^N \int_{a_n^{(k)}}^{b_n^{(k)}} \pi_{X_n}(x_n)\, \mathrm{d}x_n.
    \label{eq:normalization_pdf}
\end{equation}
\reviewag{Note that $P(Y_k=1) \neq 0$ by construction, since $a_n^{(k)}$, respectively $b_n^{(k)}$ are chosen such that $\varrho_k \neq 0$, see Section~\ref{subsubsec:refine}.}

In each element, a local interpolation grid $\mathcal{X}_{\Lambda^{(k)}}^{(k)}$ is generated following a local multi-index set $\Lambda^{(k)}$. 
The weighted Leja points in each element are obtained using the local \glspl{pdf}, where the local univariate Leja sequences are defined as 
\begin{equation}
    x^{(k)}_{n,j_n} = \argmax_{x\in d_n^{(k)}} \sqrt{\pi_{X_n}\left(x\right)/\varrho_k} \prod_{i=0}^{j_n-1} \left|x-x_{n,i}^{(k)}\right|,
\label{eq:leja_local}
\end{equation}
and correspond to local univariate interpolation grids $\mathcal{X}_{n,j_n}^{(k)} = \left\{x_{n,0}^{(k)}, x_{n,1}^{(k)}, \dots, x_{n,j_n}^{(k)}\right\} \subset d_n^{(k)}$.
Local hierarchical polynomial interpolations $\tilde{g}_k\left(\mathbf{x}^{(k)}\right) = \mathcal{I}_{\Lambda^{(k)}}[g](\mathbf{x}^{(k)})$, $\mathbf{x}^{(k)} \in d_k, k=1,\dots,K$, are then computed as described in Section~\ref{subsection:leja-colloc} and are subsequently transformed into local \glspl{pce} as discussed in Section~\ref{subsec:leja-pce}.

Once available, the local surrogate models are combined to form global surrogate models as
\begin{subequations}
\label{eq:global_surrogate}
\begin{align}
    \tilde{g}(\mathbf{x}) &= \sum_{k=1}^K \mathbb{1}_k(\mathbf{x})\tilde{g}_k(\mathbf{x}).
\end{align}
\end{subequations}
Using the \gls{pce} based global surrogate model, the mean and variance of the output can be estimated as
\begin{align}
    \mu &= \sum_{k=1}^K c^{(k)}_\mathbf{0} \varrho_k, \\
    \sigma^2 &=\sum_{k=1}^K \left(\sigma^2_k + (c^{(k)}_\mathbf{0} - \mu)^2 \right) \varrho_k,
\end{align}
where the local variances $\sigma^2_k$ are estimated as in \eqref{eq:pce-variance} \cite{wan2005adaptive}.
For uniformly distributed parameters, the local \glspl{rv} $\mathbf{X}^{(k)}$ also follow a uniform distribution which can be easily normalized to the domain $d_k$, i.e., no additional costs arise for computing the normalization factors in \eqref{eq:normalization_pdf}. 
This also allows the further use of Legendre polynomials for the \gls{pce} basis \cite{wan2005adaptive}.
If non-uniform distributions are employed, the local orthonormal polynomials must be constructed numerically \cite{wan2006multi}.

\subsection{\texorpdfstring{$hp$}{hp}-Adaptivity}
\label{sec:adaptivity}

\begin{figure}[t!]
    \centering
    \includegraphics[width=17cm]{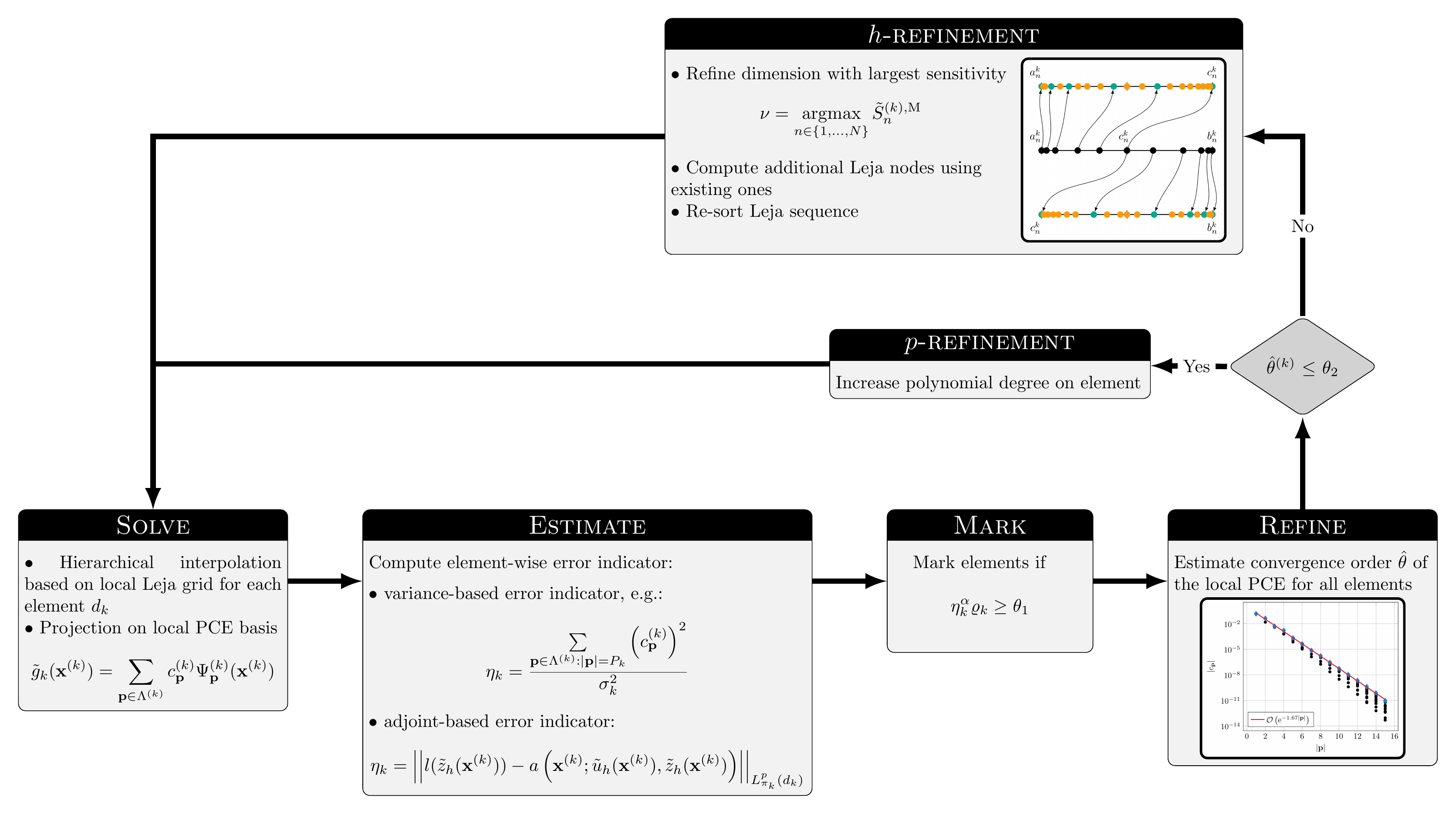}
    \caption{Sketch of the $hp$-adaptive multi-element stochastic collocation method.}
    \label{fig:flowchart}
\end{figure}

Adaptive approximation strategies based on $hp$-refinement are widely used for \gls{fem} based approximations \cite{burg2011convergence, demkowicz2006computing, dorfler2007convergence, houston2002discontinuous, mitchell2014comparison}. Typically, these strategies follow the step sequence
\begin{equation*}
    \textsc{Solve} \Rightarrow \textsc{Estimate} \Rightarrow \textsc{Mark} \Rightarrow \textsc{Refine},
    \label{eq:refinement_sequence}
\end{equation*}
which is iterated until the desired approximation accuracy has been reached.
The flowchart shown in Figure~\ref{fig:flowchart} provides a systematic overview of the $hp$-refinement procedure followed in this work for developing polynomial approximations by means of the suggested multi-element stochastic collocation method. 
The individual steps are explained in detail next.

\subsubsection{\textsc{Solve}}
Given a decomposition of the parameter (image) space $\Xi = \bigcup_{k=1}^K d_k$, a hierarchical interpolation based on the local Leja grid $\mathcal{X}_{\Lambda^{(k)}}^{(k)}$ is computed in each element $d_k$. 
Subsequently, the local hierarchical polynomial bases are transformed to orthonormal ones, thus obtaining a local \gls{pce} in each element.

\subsubsection{\textsc{Estimate}}
\label{subsubsec:estimate}
\reviewur{We consider two settings for error estimation. On the one hand, we rely on the structure of the \gls{pde} model problem of Section \ref{sec:pde_model_problem} and employ a duality-based error indicator. However, this approach requires access to the solution and matrices after finite element assembly and is therefore not always applicable. Therefore, we employ error indicators based on partial variances as a generally applicable alternative.}

\paragraph{\reviewur{Adjoint-based error estimation}} 
\reviewur{Similar to previous works\cite{jakeman2015enhancing,georg2020enhanced},  we compute the adjoint variable $z_h(\mathbf{x}) \in V_h$, subject to
\begin{equation}
    \label{eq:Helmholtz_dual}
    \reviewur{a(\mathbf{x};v_h,z_h) = q(v_h), \quad \forall v_h \in V_h}.
\end{equation}
Then, we can obtain the pointwise representation of the surrogate error 
\begin{equation}
    \label{eq:adjoint_error}
    g(\mathbf{x}) - \tilde{g}(\mathbf{x}) = l(z_h(\mathbf{x})) - a(\mathbf{x};\tilde{u}_h(\mathbf{x}),z_h(\mathbf{x})),
\end{equation}
with $\tilde{g}(\mathbf{x}) = q(\tilde{u}_h(\mathbf{x}))$. Even if \eqref{eq:adjoint_error} is not computable, because it would require knowledge of the adjoint for every parameter value, an accurate representation can be obtained by employing an adjoint surrogate $\tilde{z}_h$, which results in the error indicator 
\begin{equation}
    \label{eq:adjoint_indicator}
    \eta(\mathbf{x}) = l(\tilde{z}_h(\mathbf{x})) - a(\mathbf{x};\tilde{u}_h(\mathbf{x}),\tilde{z}_h(\mathbf{x})).
\end{equation}
Finally, we employ local error indicators 
\begin{equation}
    \eta_k = \|\eta(\mathbf{x}^{(k)})  \|_{L^p_{\pi_{k}}(d_k)} =  \left(\int_{d_k} |\eta(\mathbf{x}^{(k)})|^p \  \pi_{k}(\mathbf{x}^{(k)}) \mathrm{d} \mathbf{x}^{(k)} \right)^{1/p}, 
\end{equation}
where $p=1,2$ are common choices in our adaptive algorithm. Note that a Monte Carlo approach is used to estimate the integral, since the error indicator is cheap to evaluate.} 
\reviewag{The surrogate models $\tilde{u}_h(\mathbf{x})$ and $\tilde{z}_h(\mathbf{x})$ are created in addition to the surrogate model of the \gls{qoi}. However, they share the same multi-index sets $\Lambda^{(k)}$, equivalently, the same polynomial bases employed in $\tilde{g}(\mathbf{x})$.}

\begin{figure}[t!]
\begin{subfigure}[t]{0.48\textwidth}
    \centering
   \includegraphics[width=0.7\textwidth]{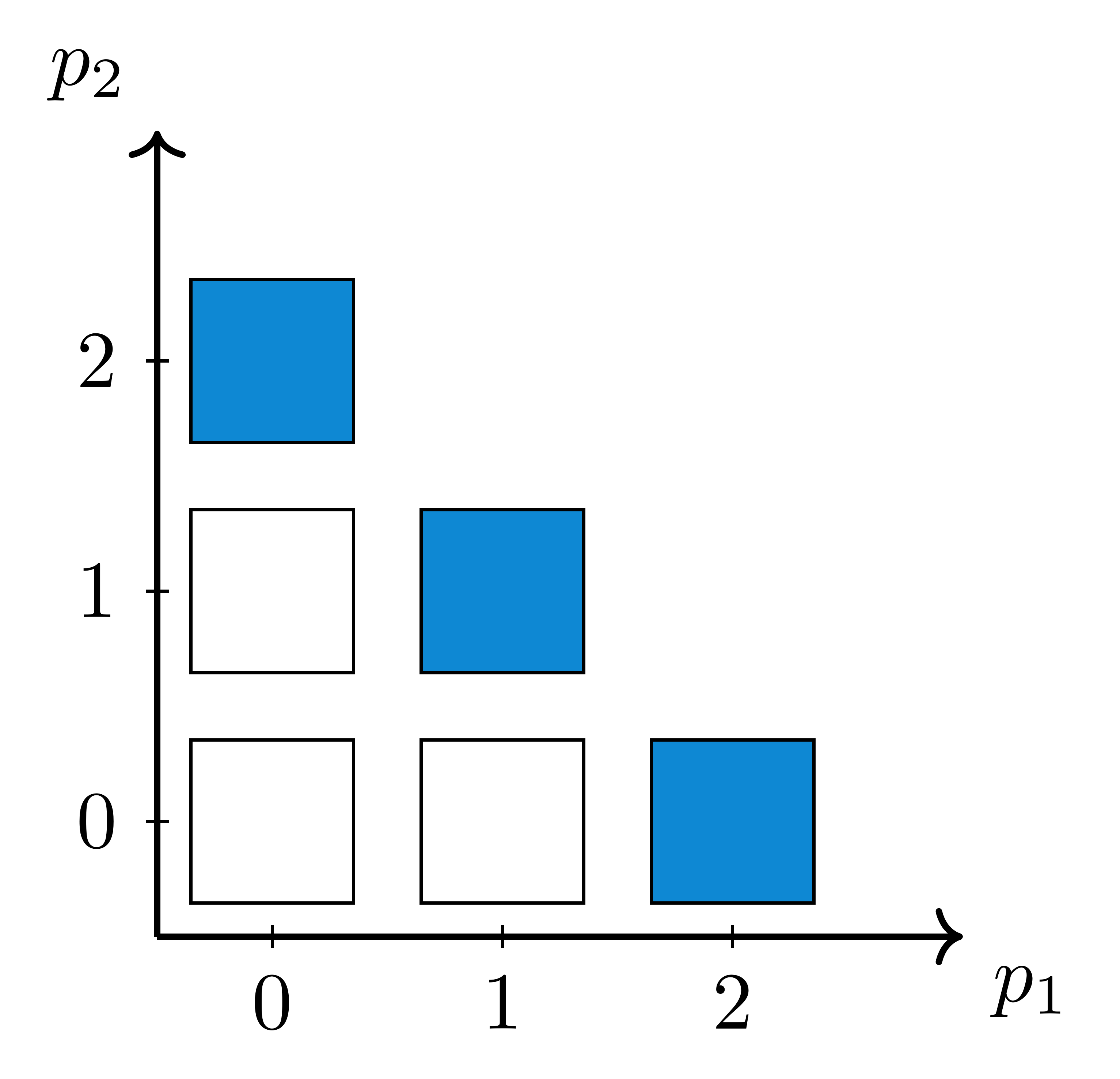} 
    \caption{}
    \label{fig:TD_index_set}
\end{subfigure}
	\hfill
	\begin{subfigure}[t]{0.48\textwidth}
	\centering
	\includegraphics[width=1.0\textwidth]{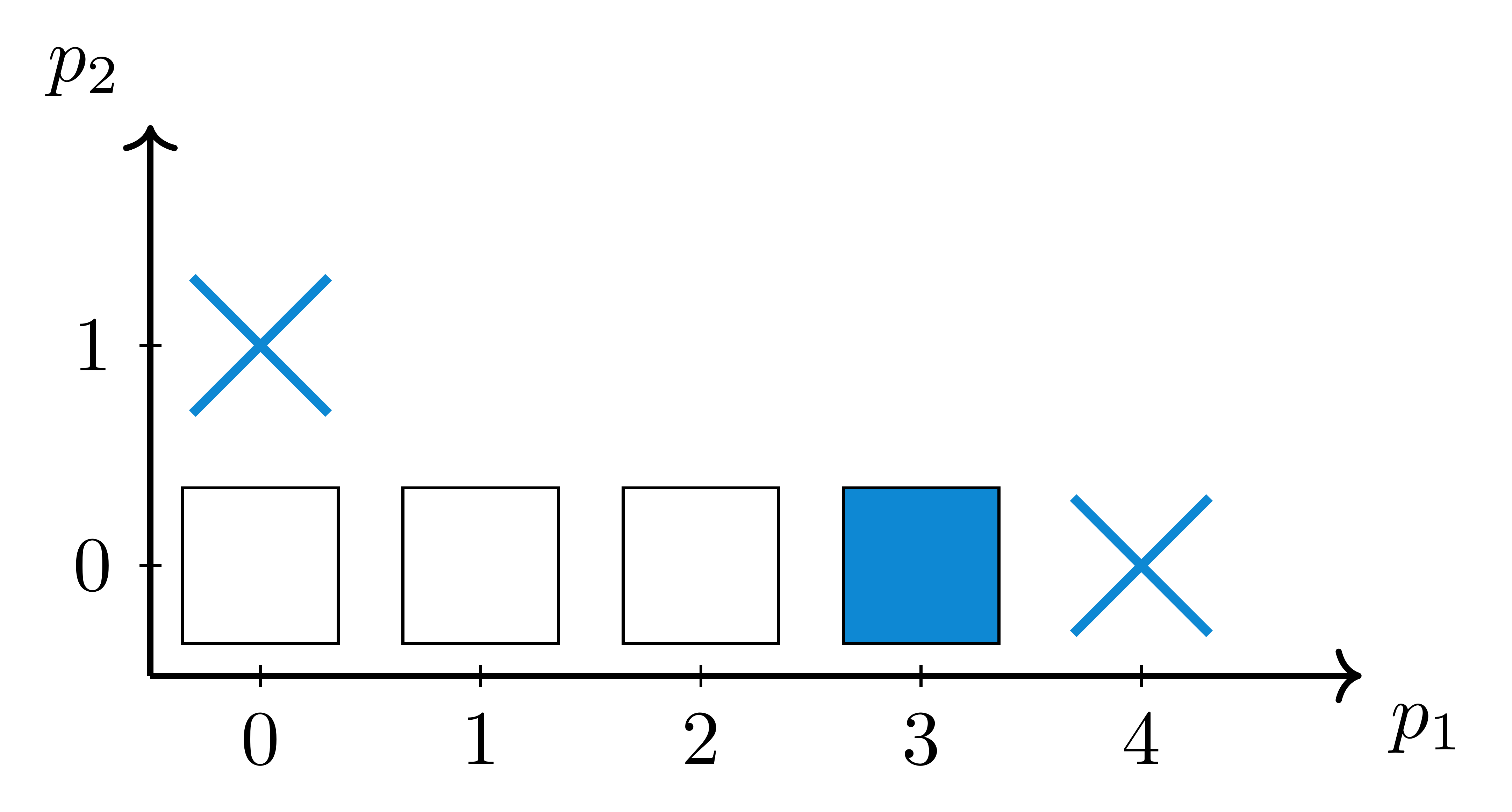} 
	\caption{}
	\label{fig:dim_adaptive_index_set}
	\end{subfigure}	
	\caption{\reviewag{Visualization of index sets for $N=2$ dimensions. The blue indices are employed in the error indicator. Squares show active indices, crosses the admissible indices. (a) Total degree basis. (b) Dimension-adaptive basis.}}
	\label{fig:index_sets}
\end{figure}

\paragraph{\reviewur{Variance-based error estimation}} 
\reviewag{Previous works have suggested error estimators for \gls{td} and Smolyak bases \cite{wan2005adaptive,foo2008multi}. In case of a \gls{td} basis, the error indicator reads}
\begin{equation}
    \sigma^{-2}_k \sum\limits_{\mathbf{p}\in\Lambda^{(k)}:\lVert \mathbf{p} \rVert_{1}=P_k}\left(c^{(k)}_\mathbf{p}\right)^{2},
  \label{eq:error_ind_wan}
\end{equation}
\reviewag{where $P_k = \max_{\mathbf{p}\in\Lambda^{(k)}}\lVert \mathbf{p} \rVert_1$.
The error indicator \eqref{eq:error_ind_wan} relates the highest-order partial variances to the total variance in an element $d_k$, see Figure~\ref{fig:TD_index_set}.}
Thus, if the estimation on the local variance $\sigma_k^2$ is sufficiently accurate, the coefficients in the nominator will be small, thus resulting in a small error indicator value as well.

In the case of an anisotropic basis, e.g., obtained with \reviewag{a dimension-adaptive algorithm, the only restriction on the local multi-index set $\Lambda^{(k)}$ is that it is downward closed}. Then, the error indicator \eqref{eq:error_ind_wan} might not be informative regarding the accuracy of the local polynomial approximation, for example, in the case of multi-index sets with a single dominant dimension. 
\reviewag{Considering the arbitrary shape of the basis, the employed coefficients may originate from different polynomial levels, unlike in \eqref{eq:error_ind_wan}. It could therefore be beneficial to employ only admissible coefficients. 
However, this can lead to cases where only a relatively small number of coefficients are employed, which might not represent the actual estimation quality of the local variance. For instance, consider an objective function for which $f(\mathbf{x}) = f(\mathbf{-x})$ and an index set as pictured in Figure~\ref{fig:dim_adaptive_index_set}. In this case, one dimension is dominant and the objective function demands for even polynomials. Hence, considering only the admissible indices, the error indicator might be misleading and not robust. 
}

\reviewag{
To solve this issue, a fixed number of coefficients is employed. Let $\Lambda^{(k)}$ denote the anisotropic multi-index set, including the admissible multi-indices. We then seek for the \gls{td} $P_k$ such that}
\begin{equation}
    \reviewag{\max_{P_k}\left\{\#\Lambda^{(k)} \ge \#\Lambda^{(k)}_\mathrm{TD} = \binom{N + P_k}{N}\right\},}
\end{equation}
\reviewag{that is, we search for a \gls{td} basis with cardinality similar to the anisotropic basis. Then, the error indicator used in the dimension-adaptive case employs the last $m_k$ added multi-indices (including the admissible indices), where $m_k$ is determined with}
\begin{equation}
\reviewag{m_k = \binom{N + P_k}{N} - \binom{N + P_k - 1}{N}. }
\label{eq:number_terms_error_ind}
\end{equation}
\reviewag{Note that with equation \eqref{eq:number_terms_error_ind} we employ as many coefficients as in the case of a \gls{td} basis. }
Assuming that $\#\Lambda^{(k)} = W_k$ and defining a nested sequence $\left\{\Lambda^{(k)}_{w_k}\right\}_{w_k=1}^{W_k}$ where $\Lambda^{(k)}_{w_k}$ is downward closed and $\#\Lambda^{(k)}_{w_k}=w_k$, the last $m_k$ multi-indices are included in the set $\Lambda^{(k)} \setminus \Lambda^{(k)}_{W_k - m_k}$. 
Then, the modified error indicator is given by 
\begin{equation}
    \reviewag{\sigma^{-2}_k \sum\limits_{\mathbf{p}\in \Lambda^{(k)} \setminus \Lambda^{(k)}_{W_k - m_k}} \left(c_\mathbf{p}^{(k)}\right)^{2}}.
    \label{eq:local_error_indicator_dim_adaptive}
\end{equation}
Due to the fact that the most recently added partial variances are employed, a similar behavior to the error indicator \eqref{eq:error_ind_wan} is to be expected. \reviewag{Note that the error-indicator \eqref{eq:local_error_indicator_dim_adaptive} is purely heuristic, however, it has proven itself a viable choice in many numerical examples, see Section~\ref{sec:numerical_experiments}.}

\subsubsection{\textsc{Mark}}
Given the local error indicators $\eta_k$, $k=1,\dots,K$, an element $d_k$ is marked for either $h$- or $p$-refinement if
\begin{equation}
    \eta_k^\alpha \varrho_k \ge \theta_1, \quad 0<\alpha<1,
    \label{eq:marking_strategy}
\end{equation}
where $\theta_1$, also referred to as the marking parameter, and $\alpha$  are user-defined constants. 
Regarding the choice of the values of these constants, an extensive discussion can be found in the literature \cite{wan2005adaptive}. 
For all numerical experiments presented in this work, the constant value $\alpha=0.5$ is used \reviewag{if a variance-based error indicator is employed and $\alpha=1$ if an adjoint-based error indicator is used. The constant} $\theta_1$ is varied according to the problem under investigation. \reviewag{A discussion on the choice of $\theta_1$ is found in Section~\ref{subsubsec:sse_function}.}
Note that marking strategies other than the criterion stated in \eqref{eq:marking_strategy} can be adopted as well, e.g., the \emph{fixed energy fraction} or the \emph{maximum strategy} criteria suggested in the literature on $hp$-adaptive \gls{fem} \cite{burg2011convergence, dorfler2007convergence}. 
In contrast to criterion \eqref{eq:marking_strategy}, these strategies mark elements if the local error indicator $\eta_k$ is large compared to the error indicators of all remaining elements. 
However, our numerical experiments showed that the proposed marking strategy delivers slightly better results. 
\reviewag{The algorithm terminates either when all local error indicators $\eta_k$ are small and/or the elements are small. Note that the local error contribution of an element to the total $L_2$ error scales with the normalization factor $\varrho_k$, which is strongly related to the element size \cite{wan2005adaptive}. Consequently, $h$-refinement around a discontinuity will eventually terminate if the elements are small enough, since the contribution to the overall error is negligible. }

\subsubsection{\textsc{Refine}}
\label{subsubsec:refine}
Once the marking procedure is complete, a decision for either $h$- or $p$-refinement for each element $d_k$ is made. 
To that end, the convergence order of the local \gls{pce} is estimated.

\begin{figure}[t!]
\begin{subfigure}[t]{0.48\textwidth}
    \centering
   \includegraphics[width=1.0\textwidth]{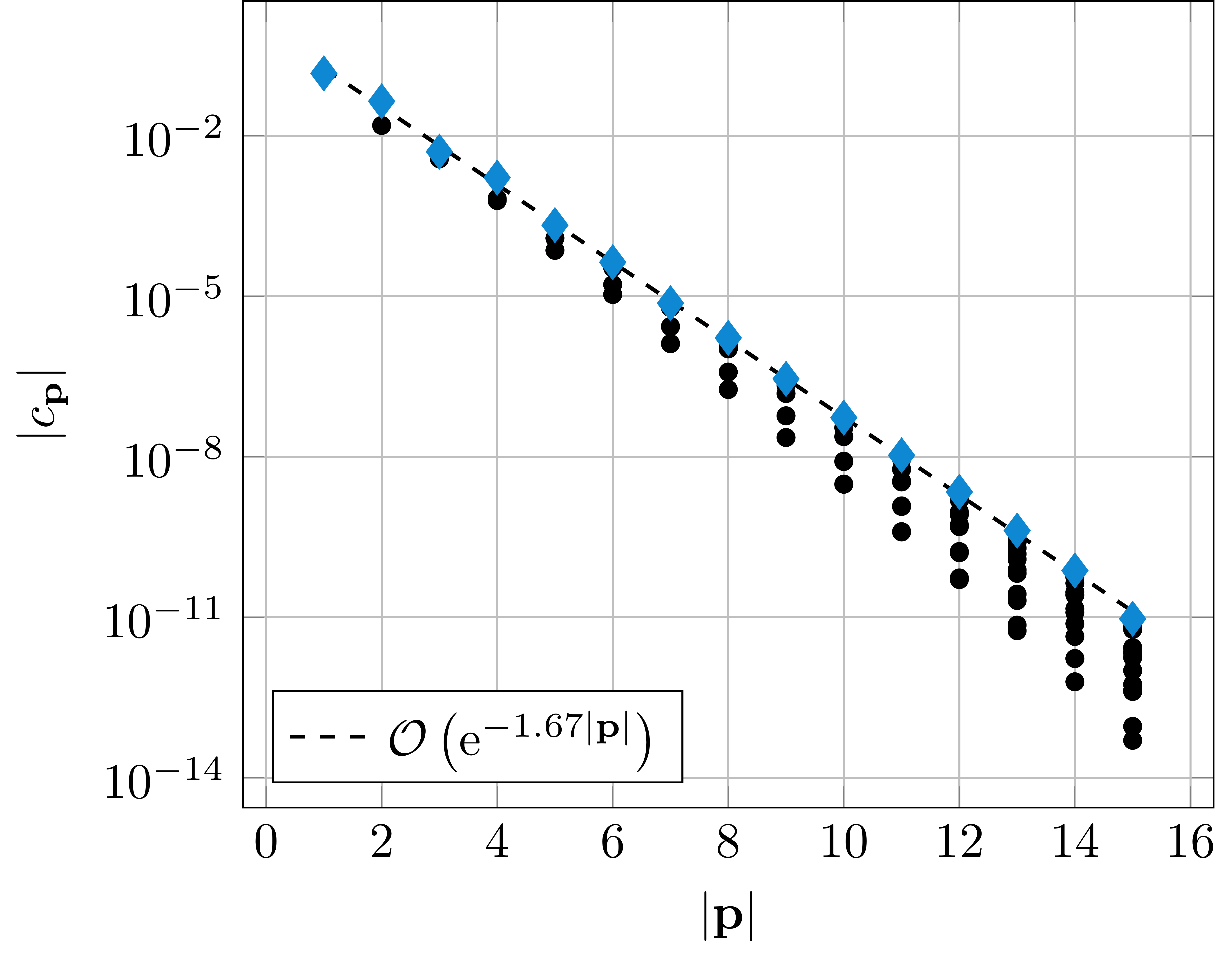} 
    \caption{}
    \label{fig:estimate_convergence_rate_smooth}
\end{subfigure}
	\hfill
	\begin{subfigure}[t]{0.48\textwidth}
	\centering
	\includegraphics[width=1.0\textwidth]{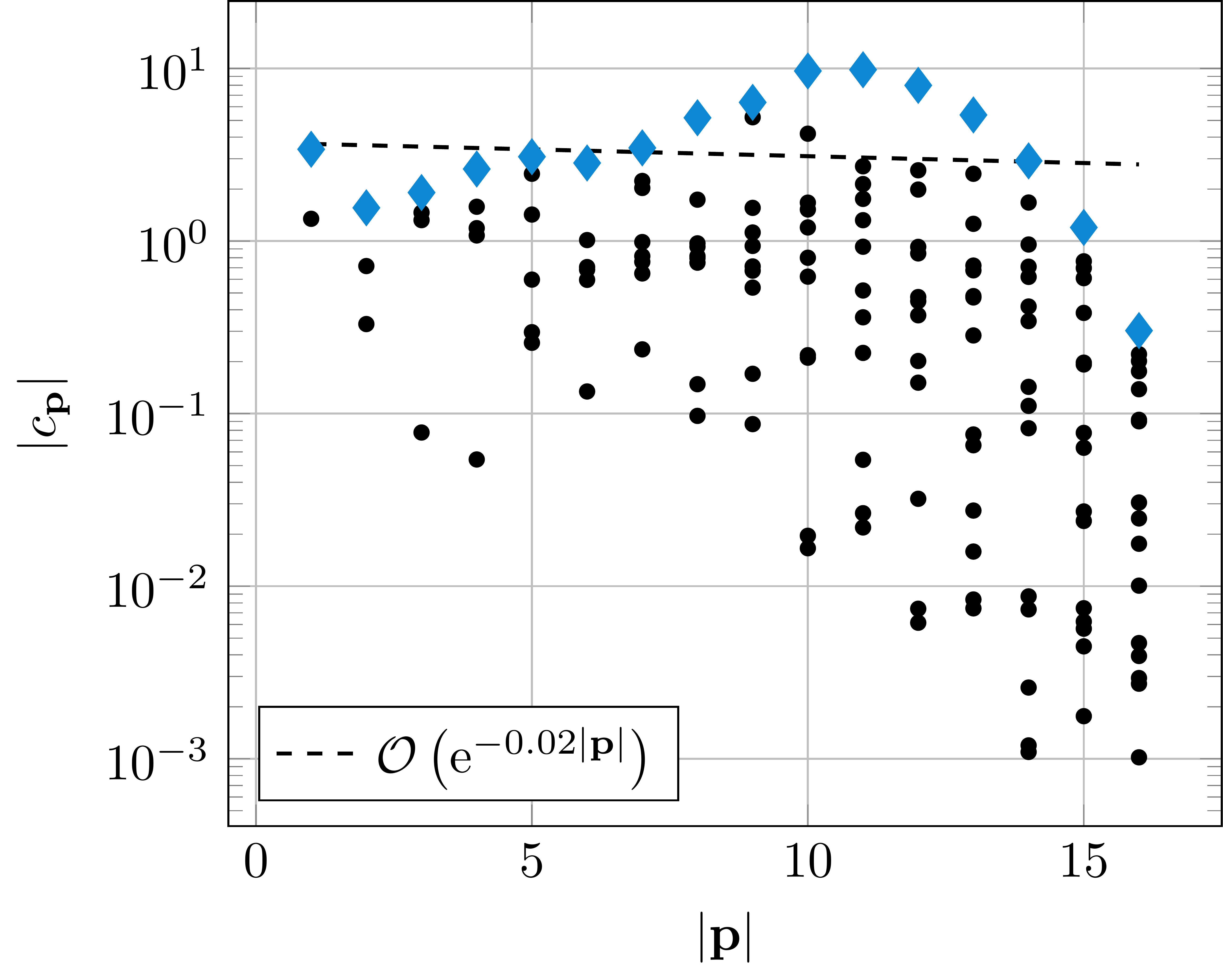} 
	\caption{}
	\label{fig:estimate_convergence_rate_reduced_reg}
	\end{subfigure}	
	\caption{Estimating the decay of the \gls{pce} coefficients, (a) for a smooth objective function  (b), for a function with reduced regularity. The blue markers show the coefficients employed to identify the rate of convergence\reviewag{, whereas the black dots show the remaining \gls{pce} coefficients.}}
	\label{fig:estimate_convergence_rate}
\end{figure}

In deterministic \gls{fe}-analysis, $p$-refinement is carried out if the solution is locally analytic on the element. In fact, the smoothness properties can be  estimated from a Legendre expansion of the solution \cite{houston2005note}. In $1$D, a linear model is fitted to the logarithmic values of the Legendre series and the slope $\hat{m}$ is extracted. Then, $p$-refinement is carried out if $\hat{\theta} = \mathrm{e}^{-\hat{m}}$ is above a certain threshold, corresponding to a sufficiently large Bernstein ellipse, hence, region of analyticity. 

A similar approach can be applied to a (local) \gls{pce} approximation 
\[
\tilde{g}^{(k)}(\mathbf{x}^{(k)}) = \sum_{\mathbf{p} \in \Lambda^{(k)}} c_\mathbf{p}^{(k)} \Psi_\mathbf{p}^{(k)}(\mathbf{x}^{(k)}).
\]
It has been shown \cite{nobile2009analysis} that for analytic functions the \gls{pce} coefficients satisfy 
\begin{equation}
    |c_\mathbf{p}^{(k)}| \le C \mathrm{e}^{- \sum_{n=1}^N \reviewur{c_n} p_n}.
\label{eq:conv_legendre_pce}
\end{equation}
In fact, the result was proven for the Legendre basis, but \reviewur{it holds for any \gls{pce} basis for which the associated density satisfies $\max_{\mathbf{x} \in \Xi}| \pi_{\mathbf{X}}(\mathbf{x})| \leq C$.} The factors \reviewur{$c_n$ depend on the size of the analytic extension of $[a_n,b_n]$ into the complex plane; the larger this extension, the larger the value of $c_n$\cite{nobile2009analysis,babuvska2010stochastic,babuska2004galerkin}}. 
In elements where the objective function $g(\mathbf{x})$ is smooth, a fast convergence, i.e., a large constant \reviewur{$c = \min_{n} c_n$} is expected, see Figure~\ref{fig:estimate_convergence_rate_smooth}. Contrarily, in elements where the objective function has less regularity, a small constant \reviewur{$c$} is expected, see Figure~\ref{fig:estimate_convergence_rate_reduced_reg}. Hence, an estimation on the convergence rate of the \gls{pce} coefficients can serve as an indicator for $p$, respectively $h$ refinement.
Hence, we define
\begin{equation}
    \hat{\theta} = \mathrm{e}^{-\reviewur{c}},
\label{eq:p_rate_estimator}
\end{equation}
as the convergence indicator. To estimate $\hat{\theta}$, a linear model with slope $\hat{m}$ is fitted to the decaying \gls{pce} coefficients. 
The convergence indicator $\hat{\theta}$ is then estimated by solving
\begin{equation}
    \begin{pmatrix}
     1 & -1 \\
     1 & -2 \\
     \vdots & \vdots \\
     1 & -P_k
    \end{pmatrix}
    \begin{pmatrix}
    b  \\
    \hat{m}
    \end{pmatrix}
    =
    \begin{pmatrix}
    \log(\max\limits_{\mathbf{p} \in \Lambda^{(k)}: \lVert \mathbf{p} \rVert_{\reviewag{1}}=1} |c_{\mathbf{p}}| ) \\
    \log(\max\limits_{\mathbf{p} \in \Lambda^{(k)}: \lVert \mathbf{p} \rVert_{\reviewag{1}}=2} |c_{\mathbf{p}}| ) \\
    \vdots \\
    \log(\max\limits_{\mathbf{p} \in \Lambda^{(k)}: \lVert \mathbf{p} \rVert_{\reviewag{1}}=P_k} |c_{\mathbf{p}}| )
    \end{pmatrix}
    \label{eq:estimate_p_conv}
\end{equation}
in the least-squares sense and determined with $\hat{\theta} = \mathrm{e}^{-\hat{m}}$. The decision for $p$- or $h$-refinement is then made as follows:
\begin{equation}
   \begin{aligned}
    p\text{-refinement for } \hat{\theta} \leq \theta_2, \\
    h\text{-refinement for } \hat{\theta} > \theta_2,
    \end{aligned} 
\end{equation}
where $\theta_2$ is a user-defined parameter. \reviewag{A discussion on the choice of $\theta_2$ is available in Section~\ref{subsubsec:sse_function}.}

\paragraph{$p$-Refinement} 
In the case of $p$-refinement, the existing polynomial basis in an element $d_k$ is extended with additional terms. 
If a \gls{td} basis is employed, the total degree  $P_k$ is increased by one order and the Leja nodes corresponding to the newly added multi-indices are added to the interpolation grid.

\reviewag{In the case of an anisotropic basis constructed with a dimension-adaptive algorithm, special treatment is necessary. The dimension-adaptive algorithm works sequentially, i.e., one node is added to the basis representation in each iteration. Hence, the entire refinement cycle, see Section~\ref{sec:adaptivity}, could be estimated after each iteration. However, the improvement of the basis representation by adding one node is expected to be only minor and is in contrast to the computational demand of the entire refinement cycle, e.g., computing the \gls{pce} coefficients. To avoid unnecessary computational costs, a larger number of nodes is added before we carry on with the next step in the refinement cycle. Following the definition of the corresponding error indicator \eqref{eq:local_error_indicator_dim_adaptive}, we estimate again the \gls{td} $P_k$ with the given cardinality $\#\Lambda$. Then, $m_k = \binom{N + P_k+1}{N} - \binom{N + P_k}{N}$ basis terms are added, i.e., the same number of basis terms as in the \gls{td} case.
Note that the number of basis terms might be slightly larger than $m_k$, since the algorithm explores all admissible multi-indices if a multi-index is added to the current set $\Lambda^{(k)}$.}

In both cases, the already existing Leja grid points along with the corresponding function evaluations are re-used, hence, the original model must be evaluated only for the newly added Leja nodes.

\paragraph{$h$-Refinement} 
In the case of $h$-refinement, for each parameter $X_n^{(k)}$, $n=1,\dots,N$, in the element $d_k$, we estimate the main-effect Sobol sensitivity index \cite{saltelli2008global, saltelli2010variance, sobol2001global} using the local \gls{pce}, such that
\begin{equation}
      \tilde{S}^{(k),\mathrm{M}}_n = \frac{1}{\sigma_k^2}\sum_{\mathbf{p}\in \Lambda^{(k),\mathrm{M}}_n} \left(c_\mathbf{p}^{(k)}\right)^{2} , \quad n=1,\dots,N,
      \label{eq:unnormalized_Sobol}
\end{equation}
where 
\begin{equation}
	 \Lambda^{(k),\mathrm{M}}_n = \left\{\mathbf{p} \in \Lambda^{(k)}: p_n \neq 0 \text{~and~} p_m=0,\,m\neq n\right\}.
\end{equation}
The parameter that corresponds to the largest sensitivity index is then selected, that is, the parameter $X_{\nu}$ with 
\begin{equation}
	\nu  = \argmax_{n \in \{1,\dots,N\}}  \tilde{S}^{(k),\mathrm{M}}_n,
	\label{eq:href_indicator}
\end{equation}
and the corresponding univariate element $d_{\nu}^{(k)} = \left[a_{\nu}^{(k)}, b_{\nu}^{(k)}\right)$ is split into a ``left'' and a ``right'' element, $d_{\nu}^{(k), \text{L}} = \left[a_{\nu}^{(k)}, c_{\nu}^{(k)}\right)$ and $d_{\nu}^{(k), \text{R}} = \left[c_{\nu}^{(k)}, b_{\nu}^{(k)}\right)$, respectively, where the value of $c_{\nu}^{(k)}$ is chosen such that the two new elements have an equal probability mass, i.e. 
\begin{equation}
\int\limits_{a_{\nu}^{(k)}}^{c_{\nu}^{(k)}} \pi_{X_{\nu}}\left(x_{\nu}^{(k)}\right) \varrho_k \,\mathrm{d}x_{\nu}^{(k)} = \frac{1}{2}.
\label{eq:h_ref_split_position}
\end{equation}
Note that if \reviewag{$b_\nu^{(k)}=b_\nu$}, then  $d_{\nu}^{(k)} = \left[a_{\nu}^{(k)}, b_{\nu}^{(k)}\right]$. 
Accordingly, the univariate interpolation grid $\mathcal{X}_{\nu, j_{\nu}}^{(k)} = \left\{x_{\nu,0}^{(k)}, x_{\nu,1}^{(k)}, \dots, x_{\nu,j_{\nu}}^{(k)}\right\}$ that exists in the univariate element $d_{\nu}^{(k)}$ is split into a left grid $\mathcal{X}_{\nu, j_{\nu_{\text{L}}}}^{(k),\text{L}}$ and a right grid $\mathcal{X}_{\nu, j_{\nu_{\text{R}}}}^{(k),\text{R}}$, such that $\mathcal{X}_{\nu, j_{\nu}}^{(k)} = \mathcal{X}_{\nu, j_{\nu_{\text{L}}}}^{(k),\text{L}} \cup \mathcal{X}_{\nu, j_{\nu_{\text{R}}}}^{(k),\text{R}}$ and $\mathcal{X}_{\nu, j_{\nu_{\text{L}}}}^{(k),\text{L}} \cap \mathcal{X}_{\nu, j_{\nu_{\text{R}}}}^{(k),\text{R}} = \emptyset$.
For $n \neq \nu$, the univariate elements $d_n^{(k)}$ and the corresponding univariate interpolation grids $\mathcal{X}_{n,j_n}^{(k)}$ remain unaffected. 
Therefore, the element $d_k$ is also split into two new elements
\begin{subequations}
\label{eq:split-d_k}
\begin{align}
d_k^{\text{L}} &= d_1^{(k)} \times d_2^{(k)} \times \cdots  d_{\nu}^{(k),\text{L}} \times \cdots \times d_N^{(k)}, \\
d_k^{\text{R}} &= d_1^{(k)} \times d_2^{(k)} \times \cdots d_{\nu}^{(k),\text{R}} \times \cdots \times d_N^{(k)}.
\end{align}
\end{subequations}
We now shift our attention to $d_{\nu}^{(k)} = d_{\nu}^{(k), \text{L}} \cup d_{\nu}^{(k), \text{R}}$, since it is the only univariate element affected by the refinement of the element $d_k$.
It can easily be observed that the interpolation grids $\mathcal{X}_{\nu,j_{\nu,\text{L}}}^{(k), \text{L}}$ and $\mathcal{X}_{\nu,j_{\nu,\text{R}}}^{(k), \text{R}}$ that result from splitting the original element, do not coincide with the Leja sequences that would be constructed in the elements $d_{\nu}^{(k), \text{L}}$ and $d_{\nu}^{(k), \text{R}}$, respectively, using the Leja sequence definition in \eqref{eq:leja_unweighted} or \eqref{eq:leja_weighted}. 
Therefore, the attractive properties of Leja sequences are mostly lost, especially their suitability as interpolation grids due to the subexponential growth of the Lebesgue constant \cite{taylor2008lebesgue}.
Nevertheless, interpolation stability can be recovered by using the existing points to initialize a Leja sequence and then add new Leja points computed as in the definitions \eqref{eq:leja_unweighted} and \eqref{eq:leja_weighted}, a procedure called \emph{stabilization} \cite{bos2015application}. 
Additionally, the stabilized interpolation grid is sorted according to the Leja ordering, which is necessary to achieve improved Lebesgue constant values \cite{bos2015application, reichel1990newton}.

\reviewur{The importance of sorting can also be understood in the case of a uniform distribution, where \eqref{eq:leja_local} is recast as 
\begin{equation}
    \label{eq:Leja_local_uniform}
    x^{(k)}_{n,j_n} = \argmax_{x\in d_n^{(k)}}  \prod_{i=0}^{j_n-1} \left|x-x_{n,i}^{(k)}\right| = \argmax_{x\in d_n^{(k)}}  | \det V(x_{n,0},\ldots,x_{n,j_n-1},x;h_n^0,\ldots,h_n^{j_n})|,
\end{equation}
that is, as determinant maximization over the Vandermonde-like matrix with points $(x_{n,0},\ldots,x_{n,j_n-1},x)$ and polynomials $\{h_n^0,\ldots,h_n^{j_n}\}.$ The connection of this greedy maximization procedure to D-optimal experimental design has already been observed \cite{narayan2014adaptive}. 
Moreover, it has been shown that the Leja-sorting procedure is equivalent to a partial pivoting LU-decomposition applied to the Vandermonde-like matrix, which can also be used to efficiently compute the Leja sequence\cite{bos2010computing}. 
}

\begin{figure}[t!]
\begin{subfigure}[t]{0.48\textwidth}
	\centering
	\includegraphics[width=1.0\textwidth]{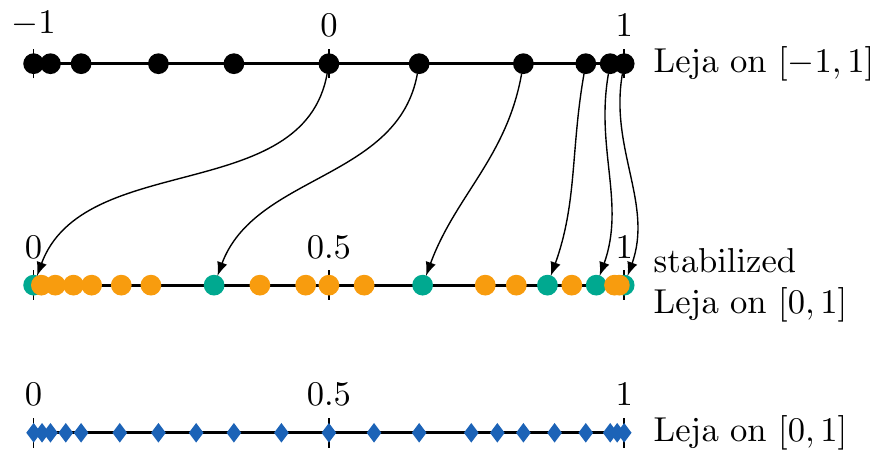} 
	\caption{}
	\label{fig:Leja_nodes_stabilized}
	\end{subfigure}	
\begin{subfigure}[t]{0.48\textwidth}
    \centering
    \begin{tikzpicture}
    \begin{axis}[grid=major, xlabel={number of interpolation points}, ylabel={Lebesgue constant}, legend style={cells={align=right}, nodes={scale=0.8, transform shape}, at={(0.65, 0.85)}, anchor=west}, legend cell align={left}]
    \addplot[thick, mark=star,TUDa-0d] table[x expr=\coordindex+1, y index=0]{figures/lebesgue_constant/lebesque_cheb_Leja_LejaSortedStab_LejaStab.txt};
    \addlegendentry{Chebyshev}
    \addplot[thick, mark=diamond*,TUDa-1b] table[x expr=\coordindex+1, y index=1]{figures/lebesgue_constant/lebesque_cheb_Leja_LejaSortedStab_LejaStab.txt}; 
    \addlegendentry{Leja}
    \addplot[thick, mark=*,TUDa-7b] table[x expr=\coordindex+1, y index=3]{figures/lebesgue_constant/lebesque_cheb_Leja_LejaSortedStab_LejaStab.txt}; 
    \addlegendentry{Leja, stabilized}
    \addplot[thick, mark=*,TUDa-3b] table[x expr=\coordindex+1, y index=0]{figures/lebesgue_constant/lebesque_LejaStabInitial.txt}; 
    \addlegendentry{Leja, initial \\ after split}
    \addplot[thick, mark=square*,TUDa-2b] table[x expr=\coordindex+1, y index=2]{figures/lebesgue_constant/lebesque_cheb_Leja_LejaSortedStab_LejaStab.txt}; 
    \addlegendentry{Leja, stabilized \\ and sorted}
    \end{axis}	
    \end{tikzpicture}
    \caption{}
    \label{fig:lebesgue_constant}
\end{subfigure}
\caption{(a) Domain splitting and stabilization of the Leja points. The black points have been calculated for a uniform \gls{rv} $X \sim \mathcal{U}[-1,1]$. The green points show the readily available interpolation nodes within $\left[0,1\right]$ after splitting the initial image space $\left[-1,1\right]$, whereas the orange points are the new points which are calculated for the uniform \gls{rv} $X \sim \mathcal{U}[0,1]$ by considering the available (green) points as initialization for the Leja sequence. For comparison, the Leja sequence calculated for $X \sim \mathcal{U}[0,1]$ without considering the readily available nodes are shown in blue. (b) Lebesgue constant for different interpolation grids.}
\label{fig:lebesgue_and_leja_nodes}
\end{figure}
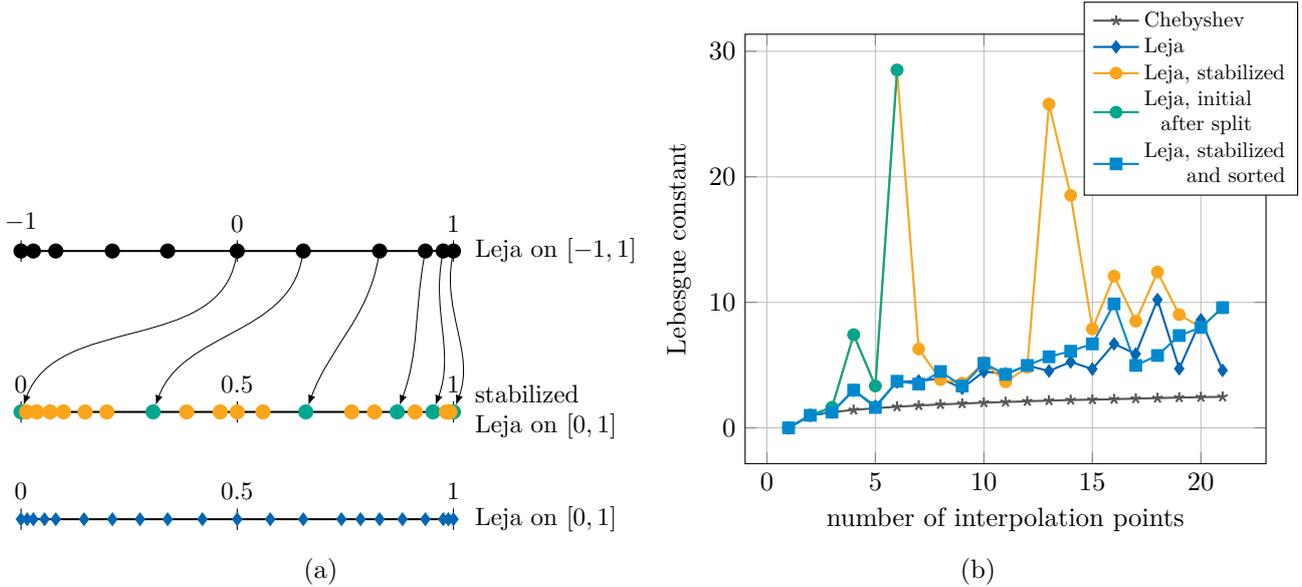

\reviewag{To the knowledge of the authors, there is no theoretical guarantee regarding the additional Leja nodes that are necessary to achieve a stable Lebesgue constant, respectively, to reduce the interpolation error to a certain level. Numerical tests in a prior work \cite{bos2015application}, as well as our own numerical results, indicate that doubling the number of existing nodes leads to a significant improvement in both Lebesgue constant and interpolation accuracy. Note that for symmetric parameter distributions, Leja points are almost symmetrically distributed. Therefore, after splitting, approximately half of the Leja points are available in each new element, see Figure~\ref{fig:Leja_nodes_stabilized}. Hence, to obtain a \gls{td} basis of the same polynomial degree, the same number of additional stabilization points must be computed. Considering that, we ensure that the number of Leja nodes is doubled for both \gls{td} and dimension-adaptive bases. Now, the multivariate interpolation grid is constructed with the desired basis representation and reads}
\begin{subequations}
\label{eq:split-d_k-grid}%
\begin{align}
\mathcal{X}_{\Lambda^{(k)}}^{(k), \mathrm{L}} &= \bigcup_{\mathbf{j} \in \Lambda^{(k)}} \left( \mathcal{X}_{1,j_1}^{(k)} \times \mathcal{X}_{2,j_2}^{(k)} \times \cdots \times \mathcal{X}_{\nu,j_{\nu_{\text{L}}}}^{(k), \mathrm{L}} \times \cdots \times \mathcal{X}_{N,j_N}^{(k)} \right),\\
\mathcal{X}_{\Lambda^{(k)}}^{(k), \mathrm{R}} &= \bigcup_{\mathbf{j} \in \Lambda^{(k)}} \left( \mathcal{X}_{1,j_1}^{(k)} \times \mathcal{X}_{2,j_2}^{(k)} \times \cdots \times \mathcal{X}_{\nu,j_{\nu_{\text{R}}}}^{(k), \mathrm{R}} \times \cdots \times \mathcal{X}_{N,j_N}^{(k)} \right).
\end{align}%
\end{subequations}%

\reviewag{Considering the case of a \gls{td} basis, all nodes are re-used by construction if we employ the multivariate grid $\mathcal{X}_{\Lambda^{(k)}}^{(k)}$ for interpolation. However, this is not true if the univariate grid $\mathcal{X}^{(k)}_{\nu,j_\nu}$ is sorted. Then, the grids $\mathcal{X}_{\Lambda^{(k)}}^{(k)} \cap \mathcal{X}_{\Lambda^{(k)}}^{(k),\mathrm{sorted}} \neq \emptyset$, but $\mathcal{X}_{\Lambda^{(k)}}^{(k)} \not\subset \mathcal{X}_{\Lambda^{(k)}}^{(k),\mathrm{sorted}}$. Depending on the sorting, several nodes and function evaluations may be lost. However, the numerical results on various examples showed that a significant amount of nodes is re-used.} 

\reviewag{
The dimension-adaptive basis is once more a special case. After $h$-refinement, the new elements most likely feature a different anisotropic behaviour than the original element. Therefore, the multi-index sets in the new elements are re-initialized with the multi-index set $\Lambda^{(k)}=\{\mathbf{0}\}$. The dimension-adaptive algorithm then utilizes the old univariate grids and builds the multivariate grid on demand. The grid defined in \eqref{eq:split-d_k-grid} then acts as a depot, where Leja nodes are picked if the adaptive Leja algorithm demands for the corresponding nodes. The univariate grids are only extended if all already existing nodes have been utilized. The different anisotropy may lead to the case where a new dimension is now dominant, which then leads to less re-use of Leja nodes. However, the dimension-adaptive basis exploits the anisotropic behaviour in the new elements and is expected to be beneficial even if not all nodes are preserved.}

The stabilization procedure and its relation to the Lebesgue constant is illustrated in Figure~\ref{fig:lebesgue_and_leja_nodes}.
We consider an initial domain $\left[-1,1\right]$ with a corresponding Leja sequence consisting of eleven points, which is split into $\left[-1,0\right)$ and $\left[0,1\right]$, and focus on the latter subdomain. 
Figure~\ref{fig:Leja_nodes_stabilized} shows the initial Leja sequence (black nodes) and the six points that are re-used within subdomain $\left[0,1\right]$ (green nodes).
The re-used points are used to initialize a Leja sequence and compute new points (orange nodes) until the interpolation grid is stabilized.
For comparison, the Leja points calculated in the interval $[0,1]$ without prior initialization are shown (blue nodes), where it is obvious that they do not coincide with the points of the stabilized Leja grid.
Figure~\ref{fig:lebesgue_constant} shows the Lebesgue constant for an increasing number of interpolation points for the stabilized grid in $[0,1]$, where the initial six points (denoted with green) are obtained from the readily available Leja sequence within $\left[-1,1\right]$ prior to splitting. 
For comparison, the Lebesgue constant for Chebyshev points and for Leja points without prior initialization within $\left[0,1\right]$ are shown.
It can easily be observed that adding new nodes according to the Leja sequence definition results in the reduction and stabilization of the Lebesgue constant.
It is also clear that sorting the stabilized grid according to the Leja ordering is crucial for interpolation stability. 

\section{Numerical experiments}
\label{sec:numerical_experiments}
In the following, the performance and approximation capabilities of the proposed $hp$-adaptive stochastic collocation method is tested on several test cases. 
A rigorous discussion on the $h$-convergence of the multi-element collocation method, in particular with isotropic and equidistant grids, is available in the literature \cite{foo2008multi}. 
Since the main advantage of using Leja sequences is the ability to re-use readily available Leja points and thus significantly reduce the computational cost due to the model evaluations on the collocation points, we focus on the computational gains provided by the $h$-adaptive and $hp$-adaptive refinement strategies described in Section~\ref{sec:me_leja}.
In particular, we first consider fixed polynomial bases, thus disregarding $p$-adaptivity, and compare the $h$-adaptive multi-element stochastic approach developed in this work against other $h$-adaptive methods found in the literature \cite{foo2008multi, wan2005adaptive, pflueger2010sgpp}. 
Unless stated otherwise, a discrete \gls{rms} error is used to evaluate the convergence behavior of the tested methods. 
The error is given by
\begin{equation}
    \epsilon_\mathrm{RMS} = \sqrt{\mathbb{E}_Q\left[\left(\tilde{g}(\mathbf{x})-g(\mathbf{x})\right)^2\right]} = \sqrt{\frac{1}{Q} \sum_{q=1}^Q\left(\tilde{g}(\mathbf{x}_q)-g(\mathbf{x}_q)\right)^2},
    \label{eq:RMS_error}
\end{equation}
which is computed using a validation sample $\{\mathbf{x}_q\}_{q=1}^Q$ drawn randomly from the joint input \gls{pdf}.

\subsection{\texorpdfstring{$h$}{h}-Refinement test cases}

\subsubsection{Benchmark problems}
\label{sec:gentz_2D}
\label{subsub:gentz2d}

In this test case, we discuss the cost savings that are obtained when re-using the Leja points in the context of $h$-refinement. \reviewag{Note that, due to the nestedness of the Leja nodes, all Leja nodes are re-used when $p$-refinement is performed. We consider the continuous and discontinuous Genz functions\cite{Genz1987}}
\begin{subequations}
\begin{align}
    \reviewag{g_1(\mathbf{x})} &\reviewag{= \exp \left( -\sum_{n=1}^N 10\cdot 2^{-n} \left| x_n - 0.5\right|\right)},  \label{eq:c_genz}\\
    \reviewag{g_2(\mathbf{x})} &\reviewag{= \left\{\begin{array}{ll} 0, &\text{if}\:\: x_1 > 0.5 \: \text{or} \: x_2 > 0.5, \\
\exp\left(\sum_{n=1}^N 10\cdot 2^{-n} x_n\right), &\mathrm{otherwise}, \end{array}\right.} \label{eq:dc_genz}
\end{align}%
\label{eq:genz_test_functions}%
\end{subequations}%
\reviewag{where the coefficients are chosen such that an anisotropic behaviour in the function response is achieved.
In the following, $x_n$ denotes realizations of the \glspl{rv} $X_n \sim \mathcal{U}[0,1]$.} 
The $h$-convergence is examined for varying values of the marking parameter $\theta_1$. 
Additionally, a comparison against the same methodology, \reviewag{i.e., also using  $h$-adaptivity only}, but based on Gauss-Legendre nodes is performed. \reviewag{In the latter case, the nodes cannot be re-used after $h$-refinement. A further comparison is performed against spatially adaptive sparse grids\cite{pflueger2010sgpp}, where the latter is based on the implementation provided by the SG++ toolbox\footnote[1]{\url{https://sgpp.sparsegrids.org/}. The adaptivty of the method is controlled through the surpluses of the hierarchical interpolation polynomials.}. 
All results habe been computed with a validation set of $Q=10^6$.}

\begin{figure}[t!] 
\begin{subfigure}[t]{0.48\textwidth}
\centering
    \includegraphics[width=1.0\textwidth]{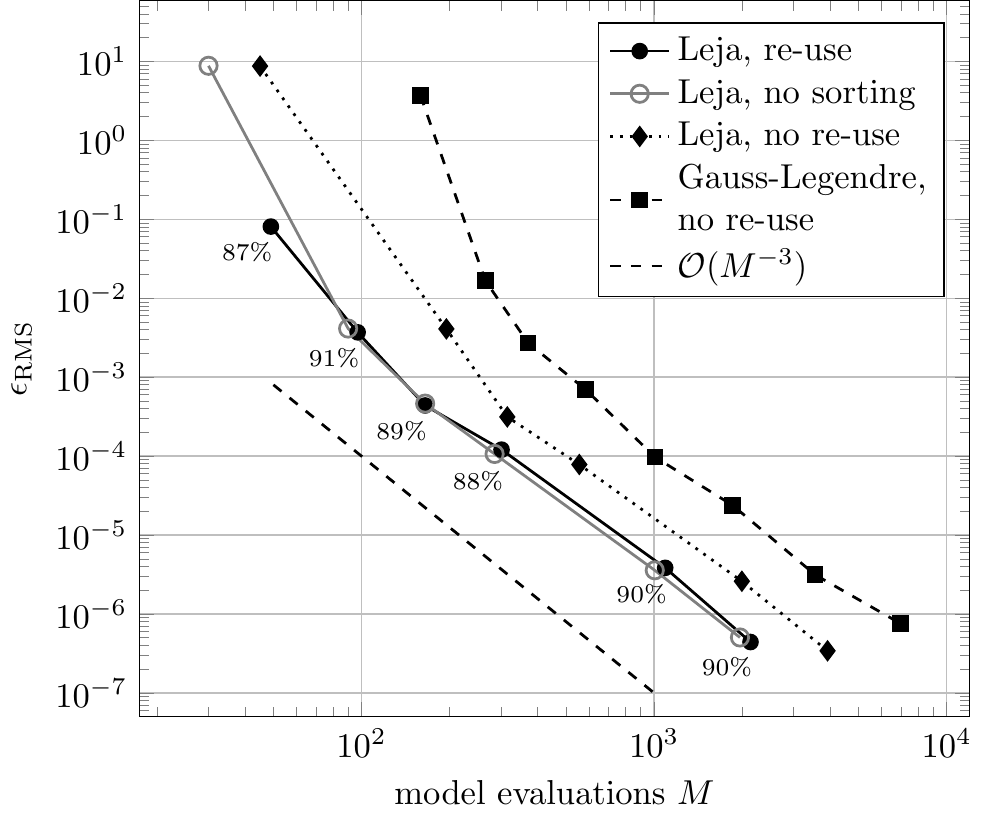}
    \caption{}
    \label{fig:Gentz_convergence}
    \end{subfigure}
	\hfill
	\begin{subfigure}[t]{0.48\textwidth}
	\includegraphics[width=1.0\textwidth]{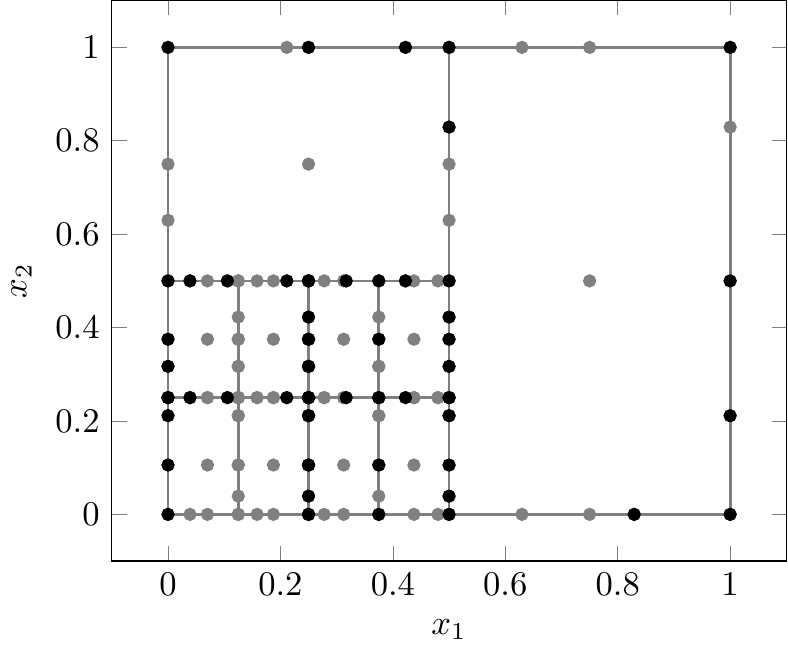}
	\caption{}
	\label{fig:Gentz_grid}
	\end{subfigure}
	\caption{(a) \gls{rms} error versus model evaluations ($M$) for the multi-element stochastic collocation method based on Leja points with and without information re-use, as well as with Gauss-Legendre nodes. \reviewag{The percentage of re-used Leja nodes is shown next to the mark.} (b) Leja interpolation grid after the convergence of the $h$-refinement procedure for $\reviewag{\theta_1=10^{-4}}$. \reviewag{The gray dots show the nodes that have been newly computed in the last iteration of the algorithm. The black dots show the nodes that have been re-used. }}
	\label{fig:Gentz_results}
\end{figure}

\reviewag{First, we consider only the discontinuous Genz function $g_2(\mathbf{x})$ for $N=2$. 
For Leja and Gauss-Legendre nodes,} a \gls{td} polynomial basis with maximum polynomial degrees $P_k=4$ in each element is used. 
Figure~\ref{fig:Gentz_convergence} shows the \gls{rms} error \eqref{eq:RMS_error} in dependence to the number of function calls after the $h$-refinement procedure has converged for a given value of the marking parameter $\theta_1$, \reviewag{with $\theta_1 \in \left\{10^{-1},10^{-3},10^{-5},10^{-6},10^{-7}\right\}$. The amount of re-used function evaluations is shown next to the marks for the case where nodes are re-used and sorted.}
As can be observed, the Leja based stochastic collocation method outperforms the Gauss-Legendre based one, irrespective of whether the Leja nodes are re-used, \reviewag{re-used but not sorted, or not re-used}.
By re-using the Leja nodes, an additional gain in performance is obtained. \reviewag{For the given example, not sorting the Leja nodes leads to a negligible improvement in the \gls{rms} error. As mentioned in Section~\ref{subsubsec:refine}, $100\%$ of the Leja nodes are re-used after element splitting if the nodes are not sorted. However, without sorting the nodes, it is unclear if the Lebesgue constant remains stable, see Figure~\ref{fig:lebesgue_constant}. Therefore, for the remaining numerical examples, the Leja nodes will always be sorted.} 

\begin{figure}[t!]
    \centering
    \includegraphics[width=1.0\textwidth]{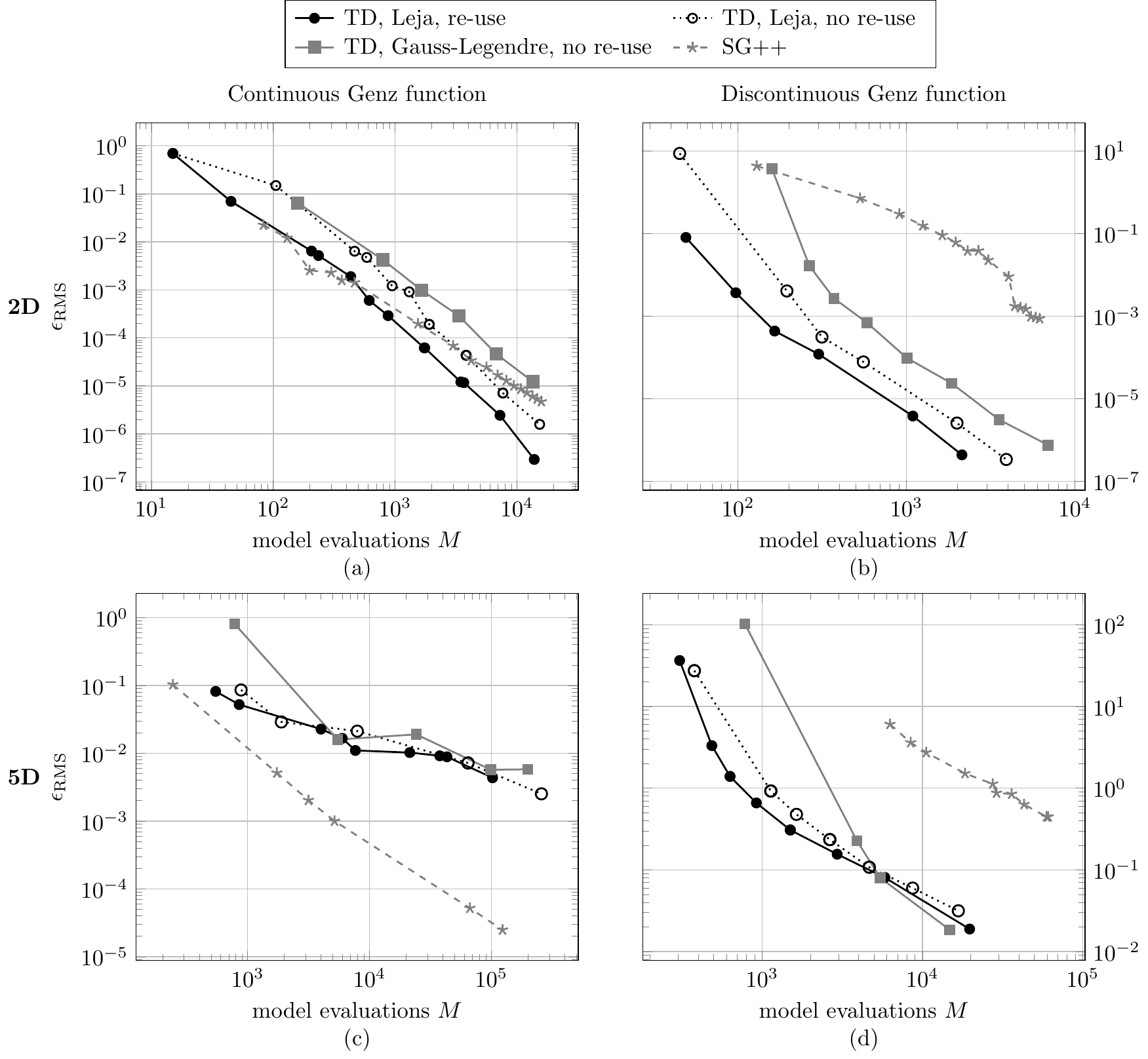}
    \caption{\reviewag{\gls{rms} error over model evaluations $M$. The results have been computed with adaptive $h$-refinement. Left column: continuous Genz function $g_1(\mathbf{x})$, right column: discontinuous Genz function $g_2(\mathbf{x})$, top row: $N=2$, bottom row: $N=5$.}}
    \label{fig:genz_h}
\end{figure}

All employed approaches reach an asymptotic $h$-convergence rate of $\mathcal{O}(M^{-3})$ once the discontinuity of the function has been isolated.
\reviewag{In the case of sorted Leja nodes, a re-use rate of approximately $90\%$ is obtained.}
Figure~\ref{fig:Gentz_grid} shows the interpolation grid and the corresponding Leja nodes after the $h$-adaptive algorithm has converged for $\reviewag{\theta_1=10^{-4}}$. \reviewag{The gray dots show the new nodes that have been computed latest by the algorithm, whereas the black dots show the nodes that have been re-used.} 
It is clearly visible that the adaptive $h$-refinement results in element splitting only in the regions of the parameter domain where the objective function is not zero, thus verifying that the $h$-refinement strategy and the corresponding refinement criteria described in Section~\ref{sec:adaptivity} perform as expected.
\reviewag{Note that, in this case, the discontinuity of the objective function coincides with the parameter axes. Therefore, the elements can resolve the discontinuity exactly. If this is not the case, a grid with dense elements around the discontinuity will appear and the convergence rate of the \gls{rms} error is hindered \cite{foo2008multi}. The algorithm stops once the criterion \eqref{eq:href_indicator} is not met. This happens if the error indicator \eqref{eq:error_ind_wan} and/or the elements are sufficiently small, equivalently, $\varrho_k$ is sufficiently small.}

\reviewag{In the following we discuss the remaining results for both test functions \eqref{eq:genz_test_functions} for $N=2$ and $N=5$. Figure~\ref{fig:genz_h} shows the \gls{rms} error over function evaluations calculated with the $h$-adaptive algorithm based on Leja and Gauss-Legendre nodes. The \gls{td} basis is constructed for a maximum polynomial degree $P_k=4$. Furthermore, results computed with the SG++ toolbox for both test functions are shown. Irrespective of the utilized nodes, the $h$-adaptive algorithm shows a similar convergence behaviour for all considered functions. The results show clearly that re-using the Leja nodes leads to a gain in the approximation accuracy. In the continuous case, the SG++ is superior for $N=5$, but inferior for $N=2$. This can be attributed to the underlying tensor product structure of the proposed multi-element approach, which is not inherent in the spatially adaptive sparse-grids approach. Contrarily, the convergence of the SG++ method is hindered when considering the discontinuous example. This behaviour is consistent with previous results \cite{pflueger2010sgpp} and is related to the regularity conditions on the objective function that are here violated.}

\subsection{\texorpdfstring{$hp$}{hp}-Refinement test cases}
 
\subsubsection{\reviewag{Introductory example}}
\label{subsubsec:sse_function}
\begin{figure}[t!]
\begin{subfigure}[t]{0.48\textwidth}
    \centering
    \begin{tikzpicture}
    \begin{axis}[grid=major,xlabel={$x$},ylabel={$g(x)$}]
    \addplot[thick, black] table[x index=0, y index=1]{figures/SEE_function/SEE_function_ref.txt};
    \end{axis}	
    \end{tikzpicture}
    \caption{}
    \label{fig:SSE_function_ref}
\end{subfigure}
	\hfill
	\begin{subfigure}[t]{0.48\textwidth}
	\centering
	\begin{tikzpicture}
    \begin{axis}[xmax=220,ymode=log,ymin=1e-14,grid=major,xlabel={model evaluations $M$},ylabel={$\reviewag{\epsilon_{\mathrm{RMS,rel}}}$}, legend style={cells={align=right}, nodes={scale=0.9, transform shape}, at={(0.01,0.27)}, anchor=west}, legend cell align={left}]
    \addplot[thick, mark=square*, gray, mark options=solid] table[x index=0, y expr=\thisrowno{1}^(1/2)]{figures/SEE_function/SEE_fcalls_relmeansq.txt};
    \addlegendentry{SSE}
    \addplot[thick, mark=*, gray] table[x index=0, y expr=\thisrowno{2}^(1/2)]{figures/SEE_function/GlobalChebyshev_fcalls_rmserror_relmeansq.txt};
    \addlegendentry{Chebyshev, global}
    \addplot[thick, mark=triangle, black] table[x index=0, y expr=\thisrowno{4}^(1/2)]{figures/SEE_function/GlobalLeja_fcalls_nelements_rmserror_expError_relmeansq.txt};
    \addlegendentry{Leja, global}
    \addplot[thick, mark=square, black] table[x index=0, y expr=\thisrowno{4}^(1/2)]{figures/SEE_function/MELeja_h_fcalls_nelements_rmserror_expError_relmeansq_pmax3_theta10.0001.txt};
    \addlegendentry{Leja, $h$-adaptive}
    \addplot[thick, mark=o, black] table[x index=0, y expr=\thisrowno{4}^(1/2)]{figures/SEE_function/MELeja_hp_fcalls_nelements_rmserror_expError_relmeansq_pmax2_thetap0.8_theta11e-12.txt};
    \addlegendentry{Leja, $hp$-adaptive}
    \addplot[thick,black,dashed] table[x index=0, y expr=\thisrowno{1}^(1/2)]{figures/SEE_function/href_asymptotic.txt};
    \addlegendentry{$\mathcal{O}(M^{-4})$}
    \addplot[thick,black,dotted] table[x index=0, y expr=\thisrowno{1}^(1/2)]{figures/SEE_function/hpref_asymptotic.txt};
    \addlegendentry{$\mathcal{O}(\mathrm{e}^{-0.32M})$}
    \end{axis}	
    \end{tikzpicture}
	\caption{}
	\label{fig:SEE_function_convergence}
	\end{subfigure}	
	\caption{(a) One-dimensional analytical function. (b) Relative mean square error $\eta$ in dependence of the number of model evaluations for SSE, global Chebyshev/Leja, and $hp$-adaptive Leja surrogates.}
	\label{fig:SEE_function_1}
\end{figure}
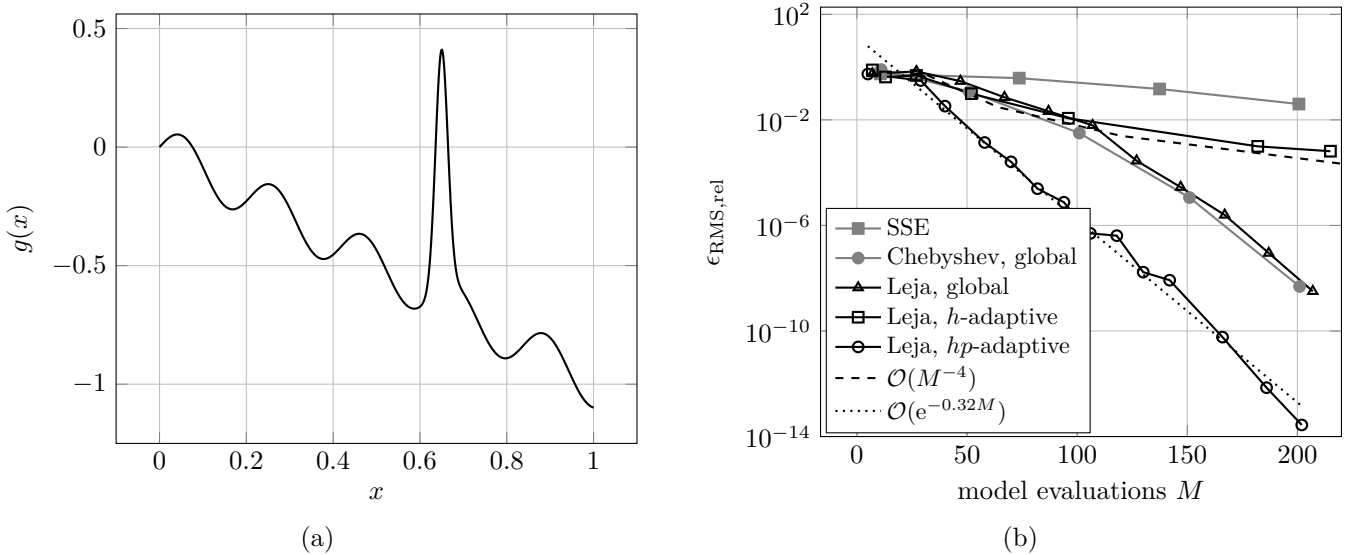

To showcase the computational benefits obtained due to the $hp$-adaptivity of the proposed method, an academic test case featuring a one-dimensional function is considered first. 
In particular, the analytical function 
\begin{equation}
    g(x) = -x + 0.1\sin(30x)+\mathrm{e}^{-(50(x-0.65))^2},
    \label{eq:sse_function}
\end{equation}
is employed, which has previously been used to validate the Spectral Stochastic Embedding (SSE) method \cite{sudret2020stochastic}. 
The function is depicted in Figure~\ref{fig:SSE_function_ref}, where the values of the parameter $x$ refer to realizations of the uniform \gls{rv} $X \sim \mathcal{U}[0,1]$. 
The observed sharp spike is the main reason why a rather slow convergence rate is to be expected if a global approximation is applied, e.g., a global interpolation on Leja or Chebychev nodes. 
The $hp$-adaptive Leja based multi-element collocation method should significantly improve the convergence rate by dividing the domain into smooth sub-domains ($h$-refinement), in which local polynomial approximations with increasing degree are developed ($p$-refinement), as presented in Section~\ref{sec:me_leja}.

Following the SSE paper\cite{sudret2020stochastic}, the accuracy of a surrogate model $\tilde{g}(x) \approx g(x)$ is evaluated using the relative root mean squared error
\begin{equation}
    \reviewag{\epsilon_{\mathrm{RMS,rel}}} = \sqrt{\frac{\mean\left[\left(g(x)-\tilde{g}(x)\right)^2\right]}{\var\left[g(x)\right]}},
    \label{eq:rmse}
\end{equation}
which is estimated using a random sample with $10^5$ points. The results for different surrogate models are shown in Figure~\ref{fig:SEE_function_convergence}. 
The results for the $hp$-adaptive Leja based multi-element collocation method are obtained with $\theta_1=10^{-12}$ and $\theta_2=0.8$.
Using $h$-refinement alone and a constant polynomial degree $P_k=3$ leads to a convergence rate of $\mathcal{O}(M^{-4})$.
For comparison, the results obtained with $p$-refined global interpolations on Leja and Chebyshev nodes, as well as with the SSE method \cite{sudret2020stochastic} are presented.

All proposed methods show a converging behavior, however, it is clearly evident that the $hp$-adaptive multi-element collocation method outperforms all other surrogate modeling options. 
Based on a least-squares estimate on the decay of the error for the $hp$-adaptive approach, a geometric convergence rate of $\mathcal{O}\left(\mathrm{e}^{-0.32M}\right)$ is obtained\cite{boyd2001}.
Note that the $hp$-adaptive Leja based multi-element stochastic collocation method and the SSE method tackle different types of problems. 
The method proposed in this work handles problems with low to moderate dimensions, whereas the SSE method is intended for high dimensional problems. 
Therefore, the comparison in Figure~\ref{fig:SEE_function_convergence} is not to be understood in a competing manner.

\begin{figure}[t!]
    \begin{subfigure}[t]{0.48\textwidth}
        \centering
        \includegraphics[width=1.0\textwidth]{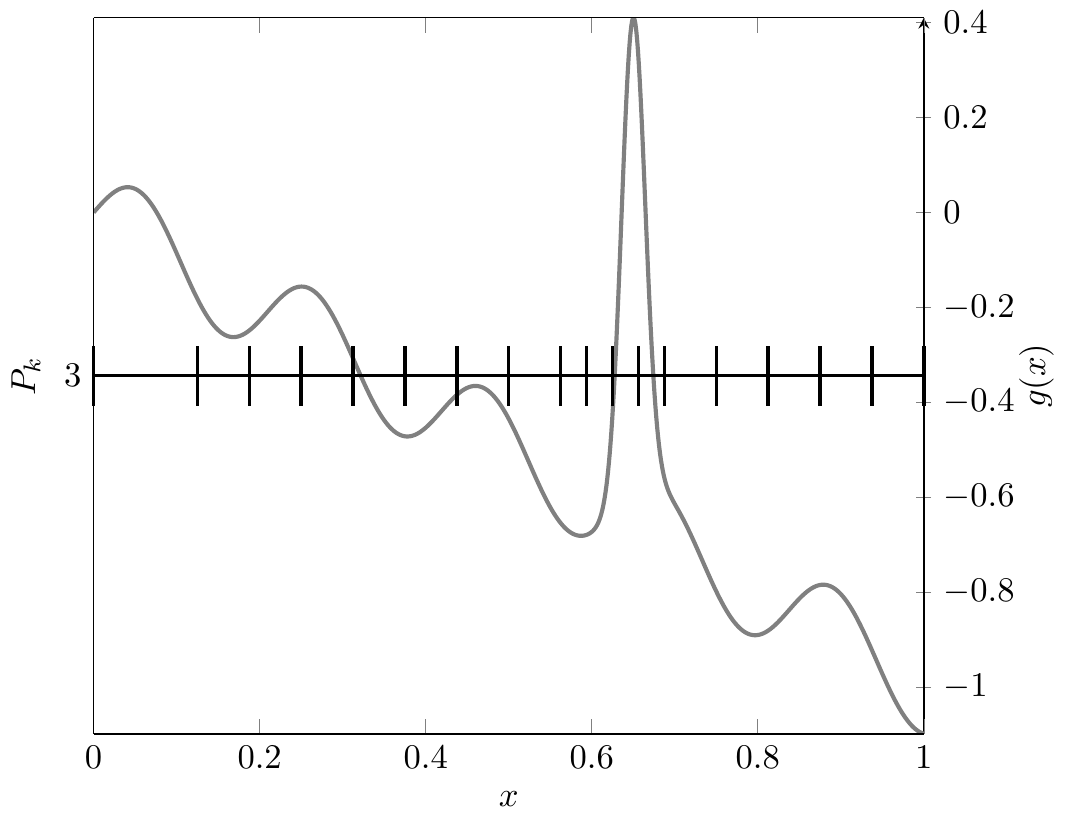} 
    \caption{}
    \label{fig:SSE_function_hp_ref_a}
    \end{subfigure}
    \hfill
    \begin{subfigure}[t]{0.48\textwidth}
        \centering
        \includegraphics[width=1.0\textwidth]{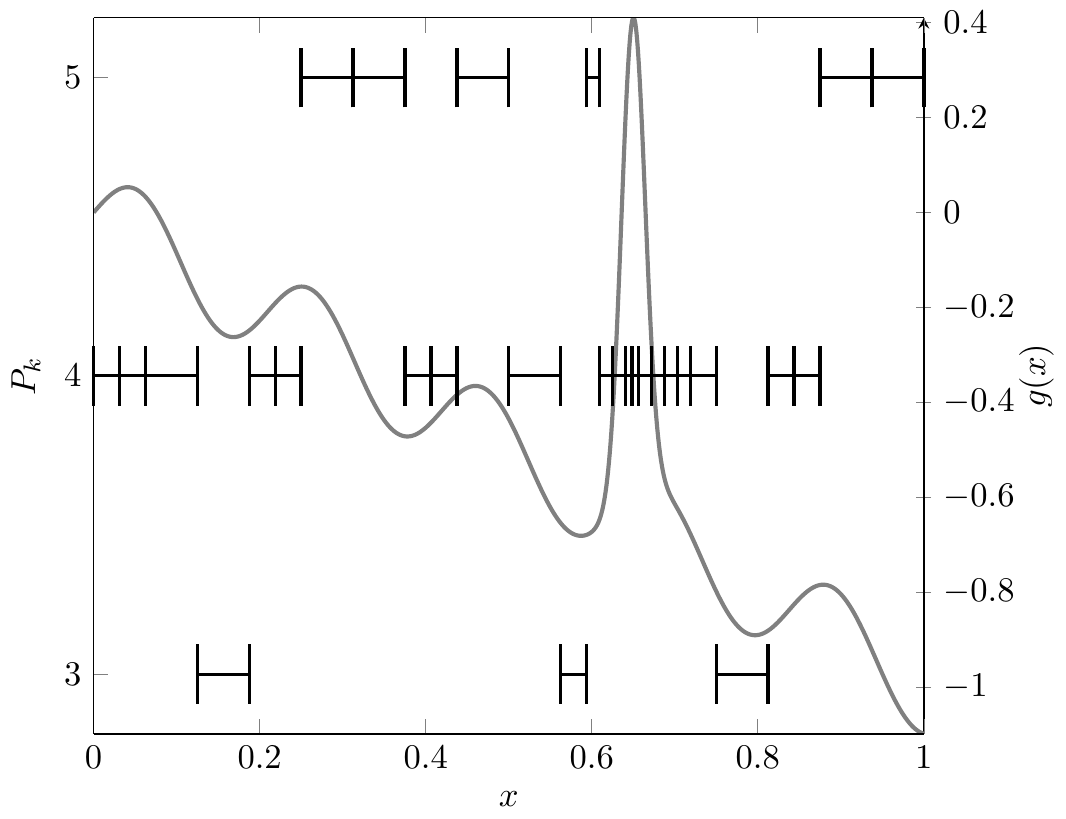} 
    \caption{}
    \label{fig:SSE_function_hp_ref_b}
    \end{subfigure}	\\
    \begin{subfigure}[t]{0.48\textwidth}
        \centering
        \includegraphics[width=1.0\textwidth]{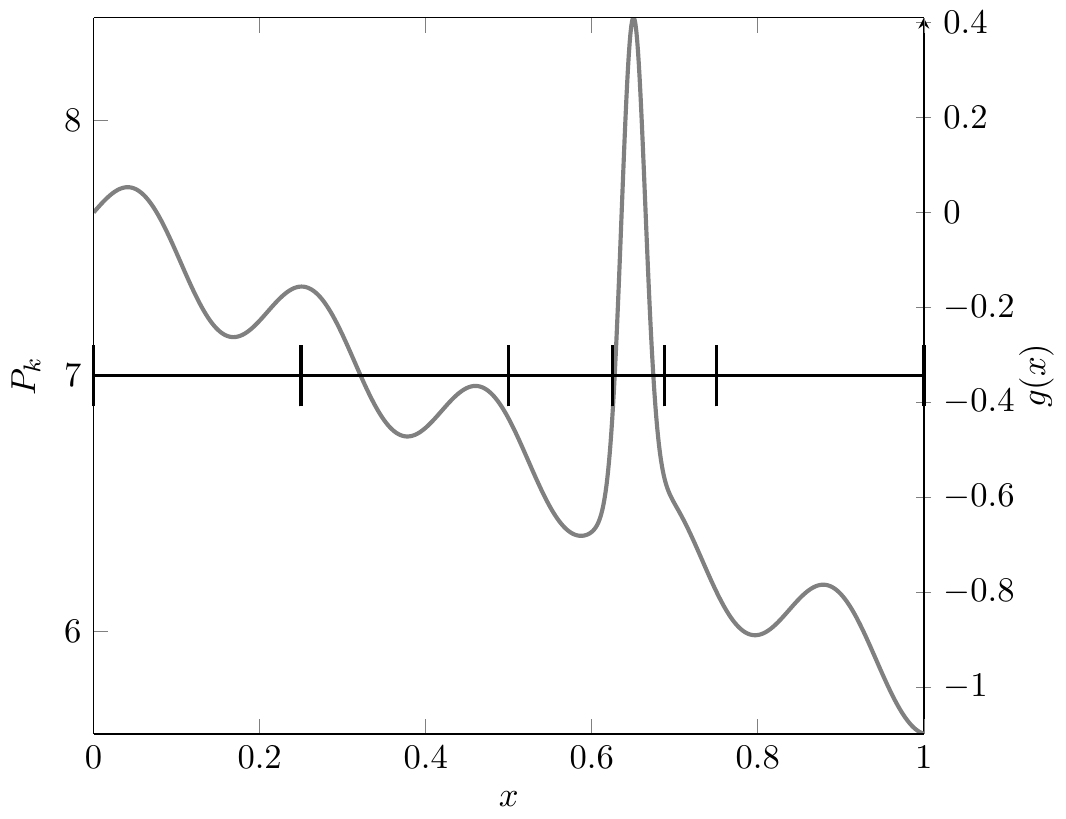} 
    \caption{}
    \label{fig:SSE_function_hp_ref_c}
    \end{subfigure}
    \hfill
    \begin{subfigure}[t]{0.48\textwidth}
        \centering
        \includegraphics[width=1.0\textwidth]{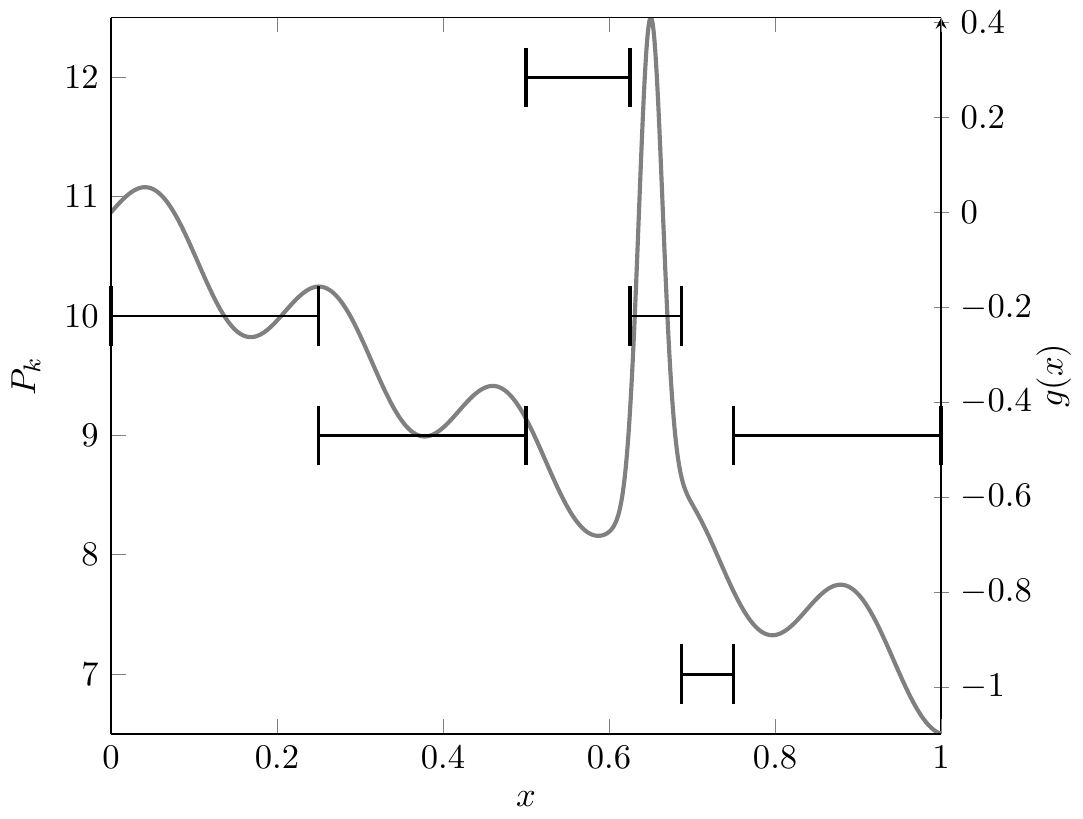} 
    \caption{}
    \label{fig:SSE_function_hp_ref_d}
    \end{subfigure}	
    \caption{Adaptively generated interpolation grids and corresponding polynomial degrees for different values of $\theta_1$ and $\theta_2$. (a) $\theta_1=10^{-2}$, $\theta_2=0.2$. (b) $\theta_1=10^{-4}$, $\theta_2=0.2$. (c) $\theta_1=10^{-2}$, $\theta_2=0.8$. (d) $\theta_1=10^{-4}$, $\theta_2=0.8$.}
    \label{fig:SEE_function_hp_ref}
\end{figure}

Next we analyze the impact of the parameters $\theta_1$ and $\theta_2$ on the $hp$-refinement procedure. 
Figure~\ref{fig:SEE_function_hp_ref} shows the domain decomposition and the polynomial degree of each element for different values of $\theta_1$ and $\theta_2$.
The left column figures correspond to $\theta_1=10^{-2}$, while the right column figures to $\theta_1=10^{-4}$.
Accordingly, the top row figures correspond to $\theta_2=0.2$, in which case $h$-refinement is preferred over $p$-refinement. 
Contrarily, $p$-refinement is prioritized in the bottom row figures, which correspond to $\theta_2=0.8$. 
An initial polynomial degree of $P_k=2$ has been used for all four cases.
All four choices for $\theta_1$ and $\theta_2$ generate interpolation grids that isolate the spike. 
As would be expected, more elements are generated for $\theta_2=0.2$, in particular in the region where the spike is observed. 
Further decreasing $\theta_1$ leads to a more finely resolved spike area, where the polynomial degrees in each element remain at moderate values. 
As would also be expected, larger elements with higher polynomial degrees are generated for $\theta_2=0.8$. 
If $\theta_1$ is further decreased in this case, $p$-refinement with a focus on the spike region is carried out. 

\reviewag{\textbf{Remark:} In the following we discuss the choice of the constants $\theta_1$ and $\theta_2$. The marking parameter $\theta_1$ determines the accuracy of the solution. However, this accuracy is estimated either with variance-based error indicators  or, if available, with the adjoint-based error indicator \eqref{eq:adjoint_indicator}, see Section~\ref{subsubsec:estimate}. Parameter $\theta_1$ is chosen according to the expected behavior of the error indicator. Numerical tests have shown that the variance-based error indicator is often too optimistic, i.e., if a desired accuracy needs to be reached, a smaller value for $\theta_1$ is necessary. In case of a fixed computational budget, $\theta_1$ can be decreased sequentially until the budget is reached. The constant $\theta_2$ is chosen according to the objective function. If we expect the function to have areas that are not easily resolvable, $h$-refinement is preferred and a smaller constant $\theta_2$ is chosen. Contrarily, if the function has areas that can easily be resolved or the function is analytic but has sharp transitions, see for example the function \eqref{eq:sse_function}, a larger $\theta_2$ should be chosen. In all considered numerical examples, the value of $\theta_2$ is chosen such that $\theta_2 \in [0,1]$. }

\subsubsection{\reviewag{Benchmark problems}}
\label{subsubsec:hp_model_problems}
\reviewag{The full capabilities of the proposed method, i.e., $hp$-adaptive refinement with different basis representations, is now discussed for the case of the Genz test functions \eqref{eq:genz_test_functions}. To assess the performance of the proposed method, we compare it against a spatially adaptive sparse grid combination technique, in the following referred to as sparseSpACE \cite{obersteiner2021_adaptiveSparseGrids}. Numerical tests are carried out for $N=2$ and $N=8$ parameter dimensions. In all cases, the error \eqref{eq:RMS_error} is determined with a validation set of size $Q=10^6$. 
}

\begin{figure}[t!]
\centering
    \includegraphics[width=1.0\textwidth]{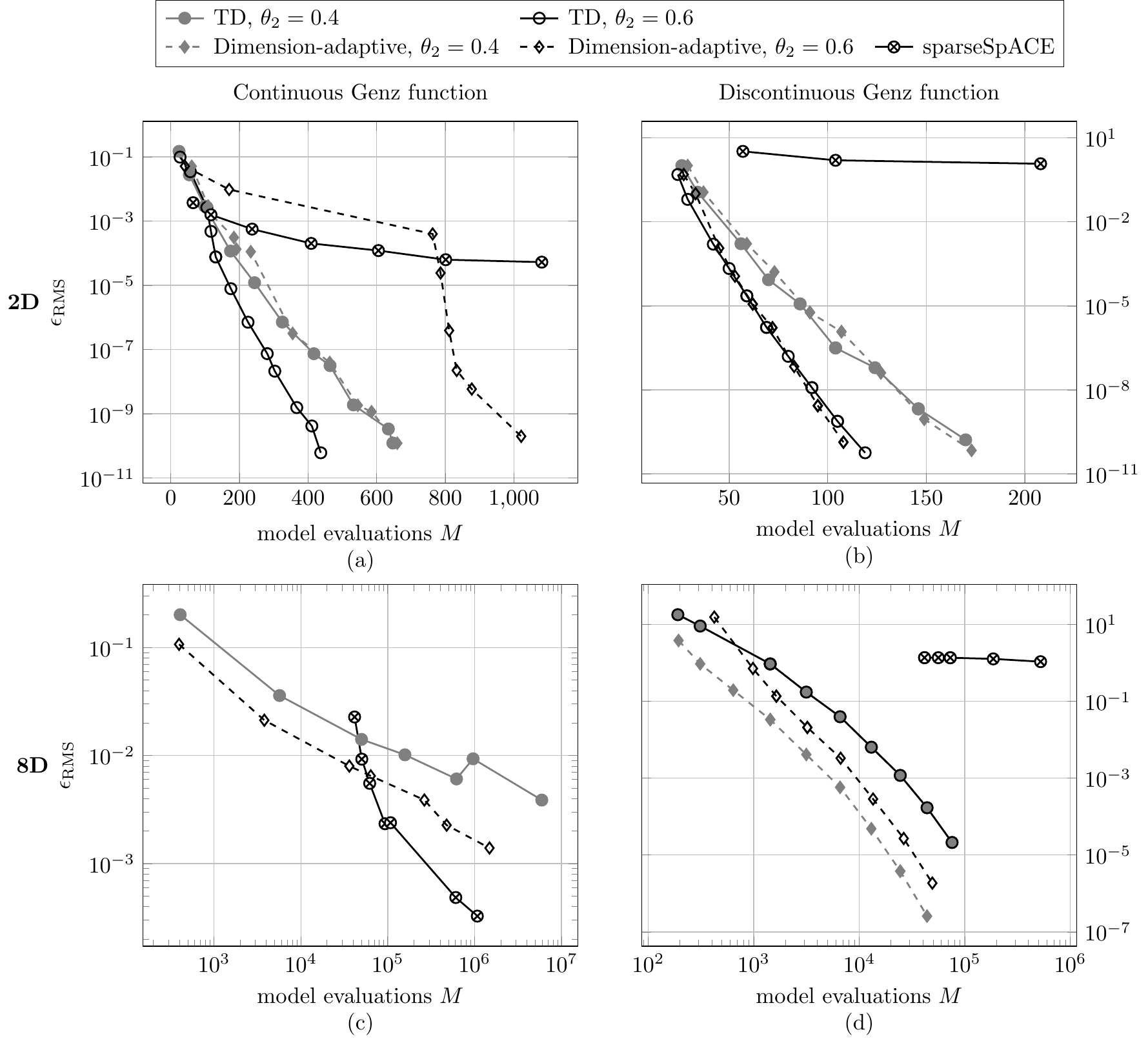}
    \caption{\reviewag{\gls{rms} error over model evaluations $M$. The results have been computed with adaptive $hp$-refinement. Left column: continuous Genz function $g_1(\mathbf{x})$, right column: discontinuous Genz function $g_2(\mathbf{x})$, top row: $N=2$, bottom row: $N=8$.}}
	\label{fig:genz_hp}
\end{figure}

\reviewag{
Figure~\ref{fig:genz_hp} shows the \gls{rms} error for both test functions for the cases $N=2$ and $N=8$. First, considering the case of $N=2$ dimensions, the $hp$-adaptive approach with a \gls{td} basis is found to be superior against all other approaches for both test functions. This result is not surprising, due to the fact that the objective functions are almost isotropic for $N=2$. The results obtained with the sparseSpACE toolbox show only a moderate convergence behaviour. Next, we consider the case of $N=8$ dimensions for the continuous Genz function, see Figure~\ref{fig:genz_hp}c. We observe that all methods show a converging behaviour. In this higher dimensional case, the dimension-adaptive basis performs significantly better than the \gls{td} basis. This can be attributed to the anisotropic objective function. Nevertheless, the tensor-product structure of the proposed multi-element approach hinders its ability to compete with \mbox{sparseSpACE} in this test case, similar to the results obtained when considering $h$-refinement only, see Section~\ref{subsub:gentz2d}.
However, looking at the results obtained for the discontinuous Genz function, see Figure~\ref{fig:genz_hp}d, the dimension-adaptive basis performs significantly better. In particular, the \gls{td} basis and the dimension-adaptive basis show a similar convergence behaviour, whereby the dimension-adaptive basis outperforms the \gls{td} basis by two orders of magnitude. In this case, the sparseSpACE toolbox does not converge, which is consistent with previously reported results \cite{obersteiner2021_adaptiveSparseGrids}.}

\reviewag{For the discontinuous Genz function \eqref{eq:dc_genz}, two $h$-refinement steps are necessary to resolve the discontinuity. All proposed $hp$-approaches achieve this minimum domain splitting for a certain parameter choice of $\theta_2$. Under this parameter configuration, the $hp$-adaptive algorithm achieves the best results. The minimum number of $h$-refinements leads to large elements within which only $p$-refinement is carried out afterwards. Then, the basis representation in the elements is of even more importance for the computational costs, respectively the accuracy of the approximation.}

\begin{table*}[b!]
\caption{\reviewag{Relative amount of re-used Leja nodes over the amount of available Leja nodes. Only savings during $h$-refinement are counted. The results were obtained upon convergence of the algorithm.}}
\centering
\ra{1.3}
\begin{tabular}{@{}llcllcll@{}}\toprule
 &  & & \multicolumn{2}{c}{Continuous Genz} & \phantom{abc}& \multicolumn{2}{c}{Discontinuous Genz} \\
\cmidrule{4-5} \cmidrule{7-8}
basis                                                                                                 & $\theta_2$ &&  $N=2$  &  $N=8$  && $N=2$ & $N=8$      \\ 
\midrule

                                                                                                             & 0.4 &&  $93\%$ & $93\%$  && $89\%$  & $38\%$   \\
\rowcolor{gray!30}[\dimexpr\tabcolsep+0.1pt\relax] \cellcolor{white}  \multirow{-2}{*}{TD}                   & 0.6 &&  $87\%$ & $88\%$  && $83\%$  & $38\%$   \\
                                                                                                             & 0.4 &&  $94\%$ & $69\%$  && $100\%$ & $78\%$   \\
\rowcolor{gray!30}[\dimexpr\tabcolsep+0.1pt\relax] \cellcolor{white}  \multirow{-2}{*}{Dimension-adaptive}   & 0.6 &&  $66\%$ & $86\%$  && $58\%$  & $50\%$   \\
\bottomrule
\end{tabular}
\label{tab:savings}
\end{table*}

\reviewag{The relative amount of re-used Leja nodes in case of $h$-refinement can be found in Table~\ref{tab:savings}. Note that the discontinuous Genz function with $N=8$ dimensions results in comparatively low re-use percentages, particularly for the \gls{td} and dimension-adaptives bases. However, in those cases, the $hp$-adaptive algorithm performed only two $h$-refinements and thus the overall costs are dominated by the following $p$-refinements.}

\subsubsection{Posterior estimation}
As stressed in Section~\ref{sec:introduction}, surrogate models for the likelihood function, respectively for the posterior distribution of inverse problems, can be an alternative to expensive Markov chain Monte Carlo analyses. 
To showcase the benefits of using the method proposed in this work, a statistical model with additive noise is considered\cite{bardsley2018computational}, which reads
\begin{equation}
    \reviewag{\mathcal{G}(\mathbf{x})} + \boldsymbol{\varepsilon} = \mathbf{b},
    \label{eq:linear_statistics_model}
\end{equation}
where $\mathbf{b}\in\mathbb{R}^D$ denotes the observations or measurements, $\mathbf{x} \in \mathbb{R}^N$ the unknown model parameters, \reviewag{$\mathcal{G}:\mathbb{R}^N \to \mathbb{R}^D$ the forward model}, and $\boldsymbol{\varepsilon} \in \mathbb{R}^D$ the Gaussian noise, such that $\boldsymbol{\varepsilon}\sim \mathcal{N}(0,\sigma^2 \mathbf{I})$, with $\mathbf{I}\in\mathbb{R}^{D\times D}$ being the identity matrix.
In the simplest case, the inverse problem \eqref{eq:linear_statistics_model} represents a linear regression problem, where $\mathcal{G}$ is the discretized model operator and $\mathbf{x}$ the vector of regression coefficients\reviewag{, accordingly, it holds that $\mathcal{G}(\mathbf{x})=\mathcal{G}\mathbf{x}$.}
In the same context, deconvolution problems pose significantly greater challenges. 

The unknown parameter vector $\mathbf{x}$ can be computed by means of \gls{mle}, such that
\begin{equation}
    \mathbf{x} = \argmax_{\mathbf{x}} L(\mathbf{x}|\mathbf{b}) = \argmax_{\mathbf{x}} \frac{1}{(2\pi)^{D/2}\sigma^D}\exp \left(-\frac{||\mathcal{G}(\mathbf{x})-\mathbf{b}||^2}{2\sigma^2}\right),
\end{equation}
where $L(\mathbf{x}|\mathbf{b})$ denotes the likelihood function.
However, inverse problems are often ill-posed, e.g. due to the amplification of the measurement noise, thus resulting in very challenging  parameter estimation tasks. 
A possible remedy is to regularize problem \eqref{eq:linear_statistics_model} by employing, e.g. Tikhonov regularization or iterative regularization\cite{calvetti2018inverse, bardsley2018computational}, to name but a few options.
Another measure is to utilize Bayesian inference \cite{calvetti2018inverse}.
Therein, the parameters $\mathbf{x}$ are modelled as \glspl{rv} and a prior distribution is assumed for them. 
Given the observations $\mathbf{b}$, information about the noise, and our prior belief regarding the parameter vector $\mathbf{x}$, Bayes' theorem yields
\begin{equation}
    \pi(\mathbf{x}|\mathbf{b}) \propto L(\mathbf{x}|\mathbf{b})p(\mathbf{x}),
\label{eq:posterior}
\end{equation}
where $p(\mathbf{x})$ denotes the prior distribution and $\pi(\mathbf{x}|\mathbf{b})$ the posterior distribution. The posterior distribution \eqref{eq:posterior} is now regularized by the prior distribution $p(\mathbf{x})$. 
The parameter vector can be estimated by maximizing \eqref{eq:posterior}, which leads to the so-called maximum a posteriori (MAP) estimation.

\begin{figure}[t!]
\begin{subfigure}[t]{0.3\textwidth}
\centering
    \includegraphics[width=1.0\textwidth]{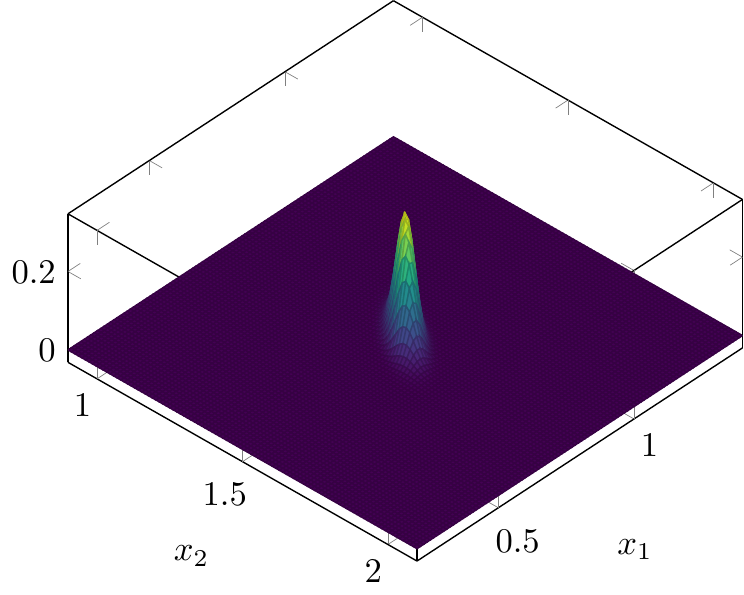}
    \caption{}
	\label{fig:bayes_2D_reference}
	\end{subfigure}
	\hfill
	\begin{subfigure}[t]{0.34\textwidth}
	\includegraphics[width=1.0\textwidth]{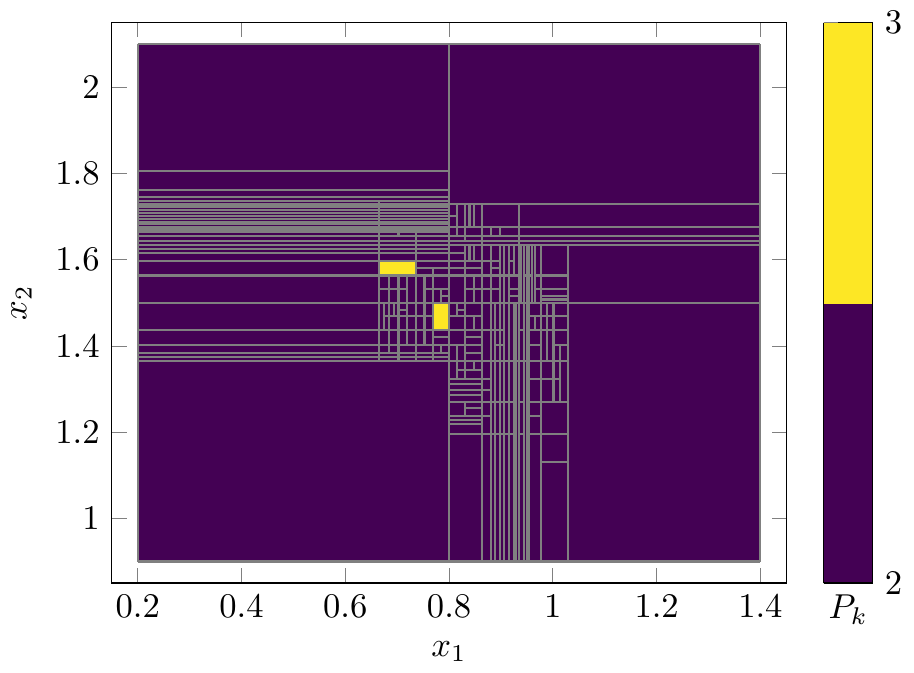}
    \caption{}
	\label{fig:bayes_2D_grid_fine}
	\end{subfigure}
	\hfill
	\begin{subfigure}[t]{0.34\textwidth}
	\includegraphics[width=1.0\textwidth]{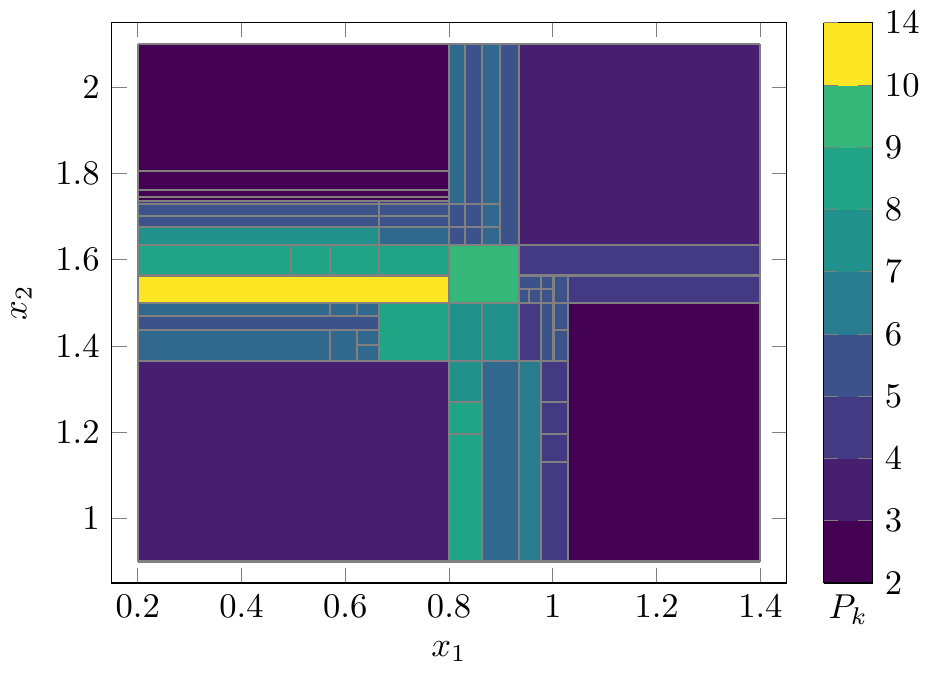}
    \caption{}
	\label{fig:bayes_2D_grid_coarse}
	\end{subfigure}
	\caption{(a) Posterior over the parameters $x_1$ and $x_2$. (b,c) Adaptively generated interpolation grids and corresponding polynomial degrees for $\theta_1=10^{-3}$ and different values of $\theta_2$. (b) $\theta_2 = 0.2$.  (c) $\theta_2 = 0.8$.}
	\label{fig:bayes_2D_results_1}
\end{figure}

Using the Bayesian approach, additional statistical quantities that characterize the posterior distribution, e.g. its mean, variance, or higher-order moments, can be computed.
In that case, the evidence, also referred to as the marginal likelihood, i.e. the normalization of \eqref{eq:posterior} over the parameter vector $\mathbf{x}$, must be computed. The computation of the evidence can be challenging, as the likelihood function and thus also the posterior can be highly concentrated, see Figure~\ref{fig:bayes_2D_reference}. The proposed multi-element collocation method can then be employed to obtain a surrogate model of \eqref{eq:posterior}, such that the evidence and the statistical moments can be computed accurately with a comparatively lower computational cost.

\paragraph{Two-dimensional regression problem}
We start with a two-dimensional regression test case where we search for the model parameters $x_1$ and $x_2$ such that 
\begin{equation}
\reviewur{	\xi \mapsto G(\xi,x_1,x_2) = x_2\exp \left(x_1 \xi\right) -2,}
\end{equation}
best fits experimental data. A measurement set with $5$ noisy observations is assumed to be available, i.e. $\left\{\xi_i,b_i\right\}_{i=1}^5$, where  $\xi_i$ are equidistantly sampled in $[-1,1]$ and \reviewur{hence $\mathcal{G}(\mathbf{x})=(G(\xi_1,x_1,x_2),\ldots,G(\xi_5,x_1,x_2))^\top$ and $\mathbf{b}=(b_1,\ldots,b_5)^\top$, respectively}. Artificial observations are then generated as $b_i = G\left(\xi_i,x_1^{\star},x_2^{\star}\right) + \varepsilon$, where $\varepsilon \sim \mathcal{N}(0,\sigma^2)$ with $\sigma^2 = 0.01$. 
Note that $x_1^{\star}$ and $x_2^{\star}$ denote the true but currently unknown model parameters.
We assume the parameters to follow a truncated normal distribution, denoted as $\mathcal{TN}(\mu,\sigma^2,\ell,u)$, where $\mu$ and $\sigma^2$ are the mean and variance of a Gaussian distribution $\mathcal{N}\left(\mu, \sigma^2\right)$ which is truncated within the interval  $\left[\ell, u\right]$. 
The parameter priors are given as $x_1 \sim \mathcal{TN}(0.8,0.04,0.2,1.4)$, $x_2 \sim \mathcal{TN}(1.5,0.04,0.9,2.1)$.

\begin{figure}[t!]
\begin{subfigure}[t]{0.48\textwidth}
\centering
    \includegraphics[width=1.0\textwidth]{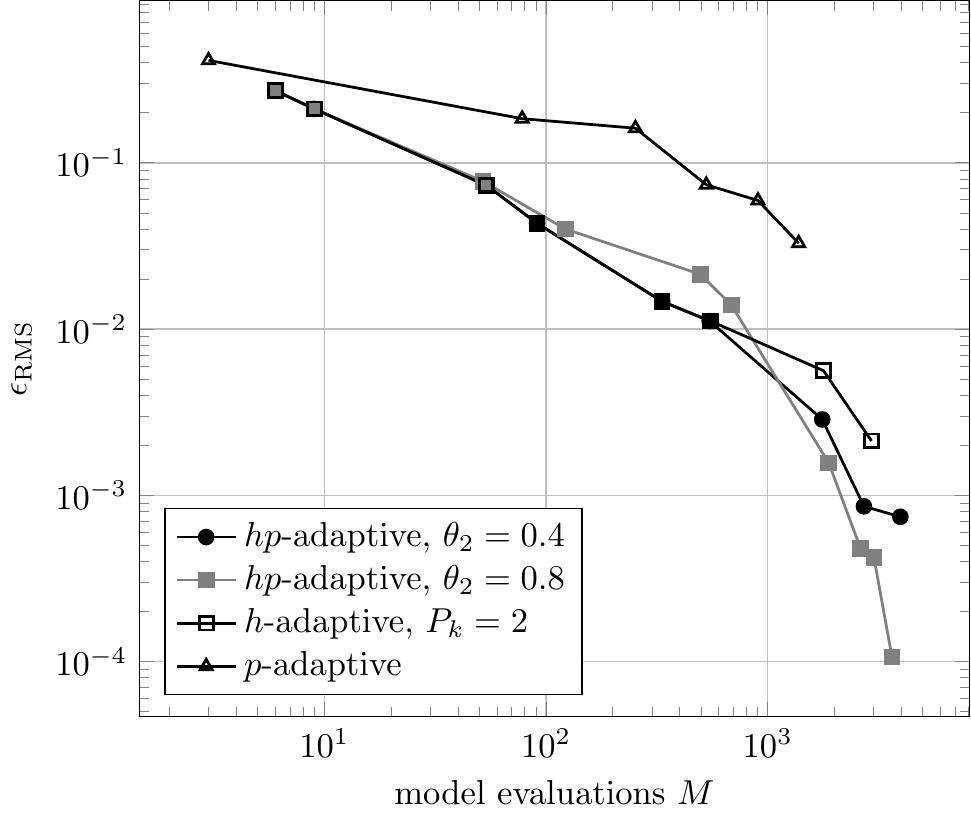}
    \caption{}
	\label{fig:bayes_2D_convergence}
	\end{subfigure}
	\hfill
	\begin{subfigure}[t]{0.48\textwidth}
	\includegraphics[width=1.0\textwidth]{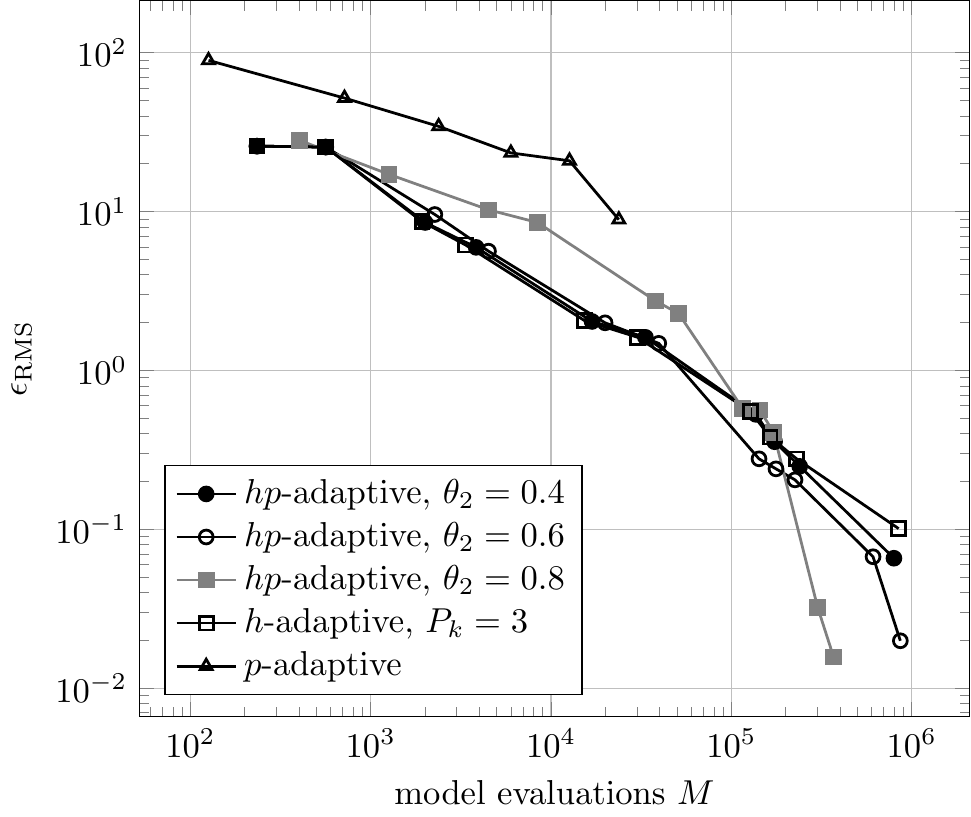}
    \caption{}
	\label{fig:bayes_4d}
	\end{subfigure}
	\caption{Convergence of the \gls{rms} error in dependence to model evaluations. The results have been obtained using TD polynomial bases. (a) Two-dimensional regression problem. (b) Four-dimensional deconvolution problem.}
	\label{fig:bayes_2D_results_2}
\end{figure}

The posterior distribution is depicted in Figure~\ref{fig:bayes_2D_reference}, which shows the strongly concentrated posterior density. Figure~\ref{fig:bayes_2D_convergence} shows the \gls{rms} error versus the number of model evaluations using stochastic collocation on Leja interpolation grids based on $h$-refinement, global interpolation with $p$-refinement, and $hp$-refinement with a varying adaptivity parameter $\theta_2$. All local polynomial approximations are based on \gls{td} bases.
Except for $p$-refinement, all marks show the results after convergence for a specific marking parameter $\theta_1$. 
The results clearly show that surrogates based on $h$- and $hp$-refinement are superior to the global collocation approach. For a moderate number of model evaluations, a similar performance is observed using either $h$- or $hp$-refinement. 
However, once the parameter domain has been sufficiently decomposed, in which case the $hp$-adaptive algorithm prefers $p$- over $h$-refinement, $hp$-adaptivity yields significant computational gains.
Additionally, at this point, a switch from an algebraic to a geometric convergence rate is observed.

Figures~\ref{fig:bayes_2D_grid_fine} and \ref{fig:bayes_2D_grid_coarse} show the domain decomposition as well as the employed polynomial degrees $P_k$ of the local \gls{td} bases, for $\theta_2=0.2$ and $\theta_2=0.8$, respectively. The results correspond to the converged algorithm for a marking parameter $\theta_1=10^{-3}$. Both figures clearly illustrate that the $hp$-adaptive approach identifies the highly concentrated posterior density. As expected, in the case of $\theta_2=0.2$, more elements are spent to isolate the posterior density, while the local polynomial degrees $P_k$ have low values. Contrarily, choosing $\theta_2=0.8$ results in fewer elements with significantly higher local polynomial degrees.  

\paragraph{Four-dimensional deconvolution problem}
We now consider a posterior estimation problem featuring four parameters. 
We again solve the inverse problem \eqref{eq:linear_statistics_model} for the parameter vector $\mathbf{x}$ and given observations $\mathbf{b}$, however, the underlying problem is now a convolution problem of the form 
\begin{equation}
    b(t) = \int_{-\infty}^{\infty} \reviewur{K}\left(t-t^\prime\right) x\left(t^\prime\right) \,\mathrm{d}t^\prime, \quad -\infty \le t \le \infty,
    \label{eq:convolution_problem_general}
\end{equation}
where $b(t)$ is the output function, $\reviewur{K}(t)$ is the weighting function or kernel, and $x(t)$ the input function. 
For example, in the field of acoustics, $b(t)$ is the system response, $\reviewag{K}(t)$ the fundamental solution of the wave equation, and $x(t)$ the \reviewur{sources}. 
The example can be found in the book of Bardsley \cite{bardsley2018computational}, where a detailed description of the functions and parameters is given. A \textsc{Matlab} implementation is also available\cite{bardsley2018github}.
The convolution problem is discretized with $4$ grid points, which leads to the discrete system \eqref{eq:linear_statistics_model}. At each of the grid points, the unknown input values $x_1,\dots,x_4$, are to be estimated based on the available data $\mathbf{b}$. Truncated normal distributions are considered as priors for the unknown input values, see Table~\ref{tab:dist_bayes4d}. Furthermore, the measurements $\mathbf{b}$ are polluted by addititive Gaussian noise with $\sigma=0.1$. The other parameters remain as in the book of Bardsley \cite{bardsley2018computational}.

Figure~\ref{fig:bayes_4d} shows the convergence of the \gls{rms} error over the number of model evaluations $M$. Similar to the two-dimensional case, all local polynomial approximations employ \gls{td} bases. As previously observed, once the concentrated posterior density has been sufficiently resolved by the algorithm, the collocation based on $hp$-refinement outperforms $h$-refinement alone. Moreover, it must be noted that, in this test case, the computational cost for constructing the Leja grid in the global $p$-refinement case, cannot be regarded to be negligible anymore due to the increasing costs of resolving the optimization problem \eqref{eq:leja_weighted}. Therefore, the $hp$-adaptive stochastic collocation is to be preferred overall. \\

\noindent \emph{\textbf{Remark:}} In both test cases presented in this section, \gls{td} bases have been employed for the polynomial approximations. The reason for this choice is that the error indicator \eqref{eq:local_error_indicator_dim_adaptive} employed for the dimension-adaptive basis expansion fails to identify the elements that need further $h$- or $p$-refinement. This issue can be attributed to the high concentration of the posterior density, in combination with the greedy algorithm of the dimension-adaptive basis approach. Nevertheless, other sparse bases such as (non-adaptive) Smolyak sparse grids \cite{smolyak1963sparsegrids, foo2008multi} can be used instead. In that case, the benefits of using Leja nodes, in particular the ability to re-use existing nodes and corresponding model evaluations, remain valid.

\begin{table}[t!]
      \caption{Distribution of the random variables for the $4$D Bayesian example.}
        \centering
	\raisebox{\depth}{\begin{tabular}[htbp]{ll}
				\toprule
				parameter & distribution \\
				\midrule
				$x_1$ & $\mathcal{TN}(0.88,0.16,-0.32, 2.08)$ \\
				$x_2$ & $\mathcal{TN}(0.1,0.04,-0.5, 0.7)$ \\
				$x_3$ & $\mathcal{TN}(0.3,0.01,0, 0.6)$ \\
				$x_4$ & $\mathcal{TN}(0.29,0.01,-0.01, 0.59)$ \\
				\bottomrule
	\end{tabular}}
	\label{tab:dist_bayes4d}
\end{table}

\subsubsection{\reviewur{Helmholtz problem}}
\label{subsubsec:Helmholtz}

\begin{figure}[t!]
    \centering
    \includegraphics[width=0.7\textwidth]{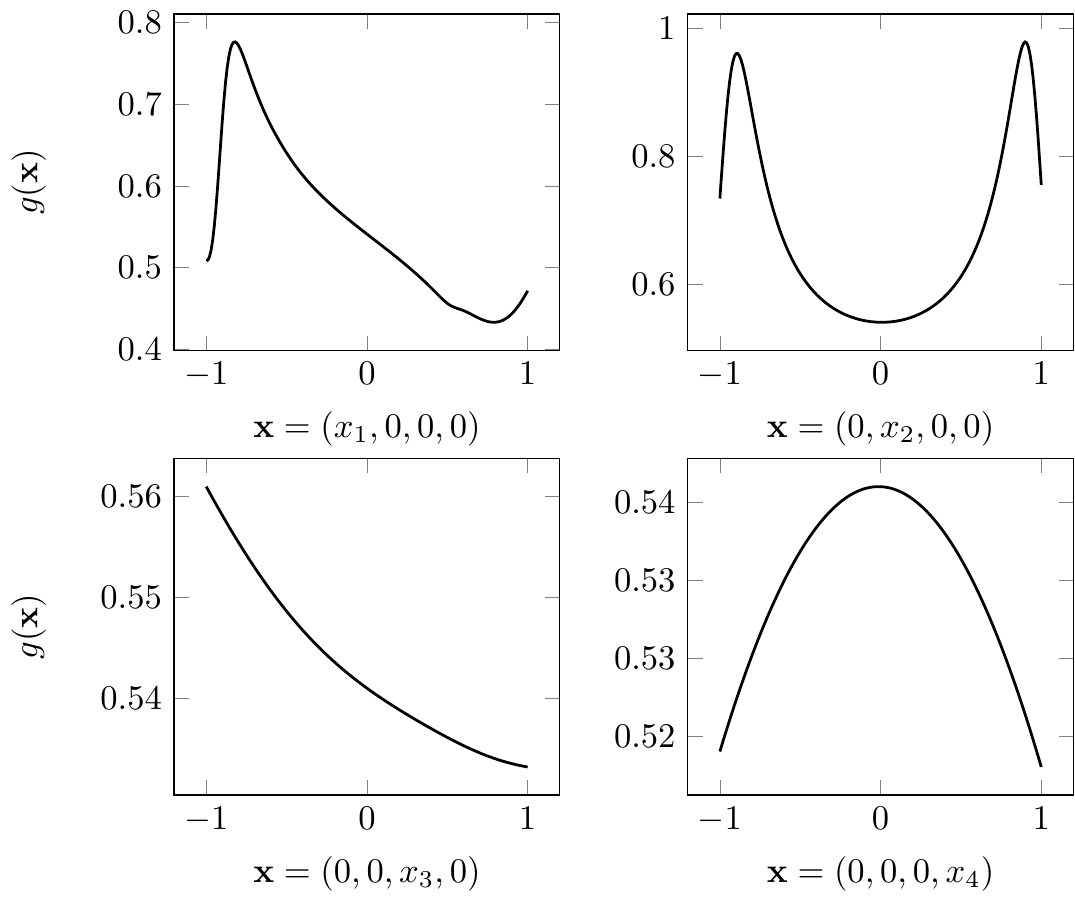}
    \caption{\reviewag{Snapshots of the quantity of interest for the Helmholtz problem for $N=4$.}}
    \label{fig:Helmholtz_4D}
\end{figure}

\begin{figure}[t!]
    \centering
    \includegraphics[width=0.5\textwidth]{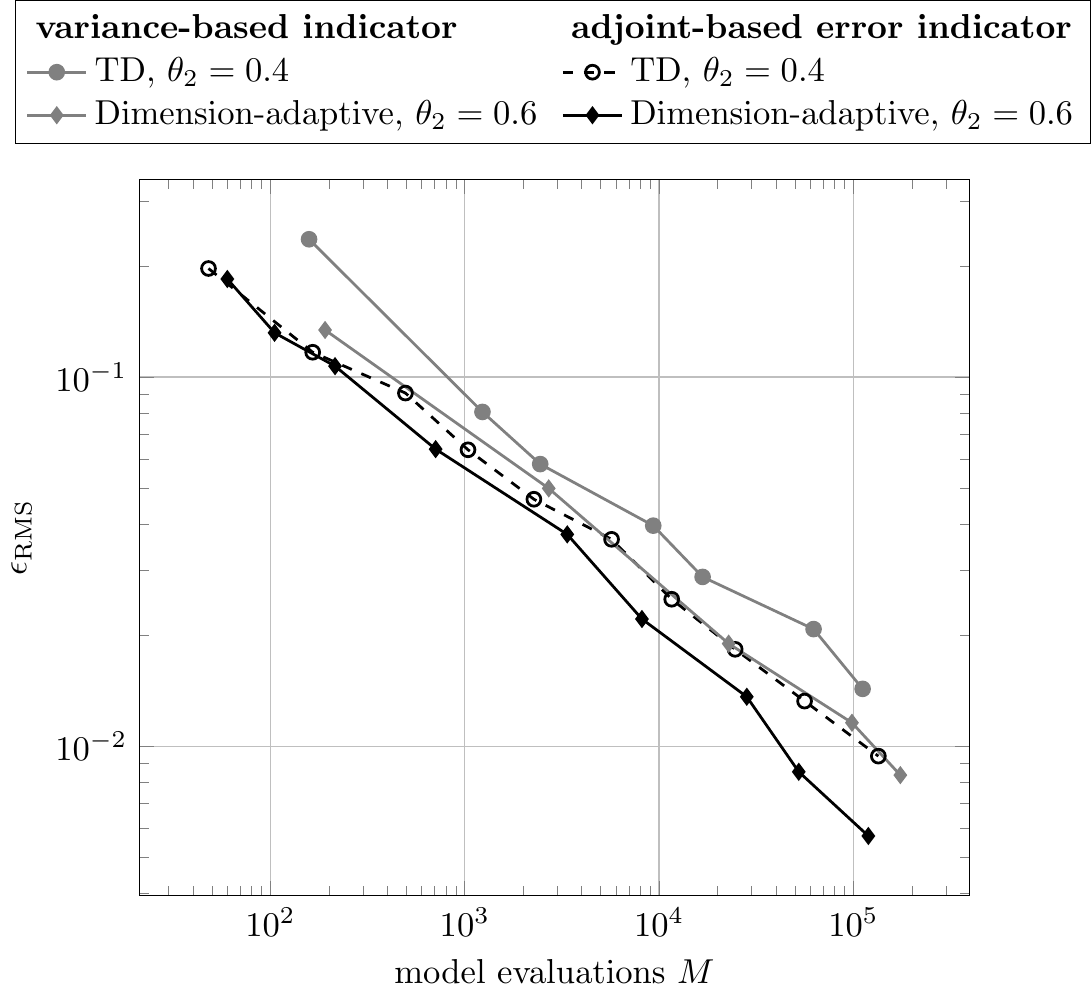}
    \caption{\reviewag{\gls{rms} error over model evaluations $M$ for the Helmholtz problem with $N=4$. }}
    \label{fig:Helmholtz_4D_convergence}
\end{figure}

\reviewur{To illustrate the method with a \gls{pde} based problem, we consider the Helmholtz equation introduced in Section~\ref{sec:pde_model_problem}.} \reviewag{For our numerical experiments on the Helmholtz equation we choose $N=4$ terms for the log-Karhunen-Lo\`eve expansion \eqref{eq:log-KLE}. One-dimensional snapshots of the \gls{qoi} are depicted in Figure~\ref{fig:Helmholtz_4D}. We assume that input parameters are distributed as $X_n\sim \mathcal{B}(2,2,-1,1),\,n=1,\dots,N$, where $\mathcal{B}(\alpha,\beta,a,b)$ denotes the Beta distribution with shape parameters $\alpha,\,\beta$ and image space $[a,b]$. The \gls{rms} error is computed with a validation set of $Q=10^5$. Figure~\ref{fig:Helmholtz_4D_convergence} compares the results that are obtained for the variance-based error indicator and the adjoint-based error indicator. The adjoint-based error indicator behaves as expected and provides a better approximation of the actual error. Consequently a higher accuracy for a fixed computational budget is achieved if the adjoint-based error indicator is utilized.}

\section{Conclusion}
\label{sec:conclusion}
In this work, an $hp$-adaptive multi-element collocation method based on Leja nodes has been introduced. The presented method allows to re-use already computed model evaluations after $h$- or $p$-refinement. The local interpolation grids are stabilized after $h$-refinement by adding additional nodes and appropriately reordering the Leja sequence. To tackle the computationally demanding case of higher dimensional problems, a dimension-adaptive basis expansion algorithm has been employed. The work considered uniform, \reviewag{truncated normal, and beta} distributed random inputs.

The numerical results showed that in the case of $h$-refinement, \reviewag{the proposed method is superior against the original multi-element approaches that do not utilize information re-use. For low dimensions and functions with discontinuities, the $h$-adaptive multi-element approach is the method of choice when compared against the spatially adaptive sparse-grids method. In higher dimensional settings with continuous functions, approaches based on spatially adaptive sparse grids are more efficient, because of the limitations of the underlying hypercube structure in higher dimensions. Furthermore,} the results also indicated that $hp$-adaptive refinement is to be preferred over pure $h$-refinement, \reviewdl{especially when higher-dimensional problems are considered}. \reviewag{In case of $hp$-refinement, the proposed method was compared to the generalized spatially adaptive sparse-grids combination technique, which is inferior if the objective function features discontinuities or features low parameter dimensions. Again, the opposite is true for higher dimensional problems with continuous functions}. Moreover, the proposed method was capable of approximating a highly concentrated posterior density in a two- and a four-dimensional inverse problem. \reviewag{Finally, the proposed method was applied to a \gls{pde} example (Helmholtz equation), where we showed that a deterministic, adjoint-based error indicator is to be preferred, if available.}

In summary, we may conclude that the proposed $hp$-adaptive collocation method based on Leja nodes is a reliable approach for surrogate modelling and \gls{uq} for functions with reduced regularity, as well as for functions featuring large gradients or sharp transitions.
\reviewag{We expect, that the computational demand of the proposed method can be further reduced when combined with methods that detect the critical areas in the parameter domain beforehand \cite{jakeman2013minimal, kawai2020multi, kawai2022gegenbauer}. Then, a reduced number of $h$-refinement steps is expected, which in turn should reduce the overall complexity.}

\subsection*{Acknowledgment}
D. Loukrezis and H. De Gersem would like to acknowledge the support of the Graduate School of Excellence for Computational Engineering at the Technische Universit\"at Darmstadt. D. Loukrezis further acknowledges the support of the DFG via the research contract 361 (project number: 492661287). A. Galetzka's work is supported by the DFG through the Graduiertenkolleg 2128 ``Accelerator Science and Technology for Energy Recovery Linacs''.

\bibliographystyle{abbrv}


\end{document}